\input harvmac.tex

%\draftmode

%=================================================================

\input amssym.def \def\Z{{\Bbb Z}}  \def\Q{{\Bbb Q}}
  \def\N{{\Bbb N}}

\def\frac#1#2{{\textstyle{#1\over #2}}}

% standard tableaux
\def\b#1{\kern-0.25pt\vbox{\hrule height 0.2pt\hbox{\vrule
width 0.2pt \kern2pt\vbox{\kern2pt \hbox{#1}\kern2pt}\kern2pt\vrule
width 0.2pt}\hrule height 0.2pt}}
\def\ST#1{\matrix{\vbox{#1}}}
\def\STrow#1{\hbox{#1}\kern-1.35pt}
\def\bv{\b{\phantom{1}}}

% triangles
\def\tri#1#2#3#4#5#6#7#8#9{\matrix{#4\cr
	#3\quad#5\cr #2~\qquad #6\cr #1\quad #9\quad#8\quad#7\cr}}

%troisieme type de triangles

%triangles plus gros

%triangles plus gros#2

% alignements multiples
\def\eqalignD#1{
\vcenter{\openup1\jot\halign{
\hfil$\displaystyle{##}$~&
$\displaystyle{##}$\hfil~&
$\displaystyle{##}$\hfil\cr
#1}}
}
\def\eqalignT#1{
\vcenter{\openup1\jot\halign{
\hfil$\displaystyle{##}$~&
$\displaystyle{##}$\hfil~&
$\displaystyle{##}$\hfil~&
$\displaystyle{##}$\hfil\cr
#1}}
}

\def\text#1{\quad\hbox{#1}\quad}

\def\la{\lambda}

\def\nuh{{\hat \nu}}
\def\muh{{\hat \mu}}

\def\lah{{\hat \lambda}}

\def\om{{\omega}}
\def\y{{\infty}}

\def\x{{x^{-1}}}

\def\rw{\rightarrow}

\def\Om{\mathop{\Omega}\limits}
\def\O{{\displaystyle\Om_{\geq}}}
\def\Ox{{\displaystyle\Om_{\geq}^x}}
\def\Oy{{\displaystyle\Om_{\geq}^y}}
\def\Os{{\displaystyle\Om_{\geq}^s}}
\def\Ot{{\displaystyle\Om_{\geq}^t}}

\def\Oxx{{\displaystyle\Om_{=}^x}}
\def\OR{{\displaystyle\Om_{=}^R}}
\def\OQ{{\displaystyle\Om_{=}^Q}}
\def\ON{{\displaystyle\Om_{=}^N}}
\def\OM{{\displaystyle\Om_{\geq}^M}}

\def\Nc{{\cal N}}
\def\Rc{{\cal R}}

\overfullrule=0pt

% Equations (overrides harvmac's equation macros)
\newcount\eqnum  
\eqnum=0
\def\eq{\eqno(\secsym\the\meqno)\global\advance\meqno by1}
\def\eqlabel#1{{\xdef#1{\secsym\the\meqno}}\eq }  

% References (overrides harvmac's reference macros)
\newwrite\refs 
\def\startreferences{
 \immediate\openout\refs=references
 \immediate\write\refs{\baselineskip=14pt \parindent=16pt \parskip=2pt}
}
\startreferences

\refno=0
\def\aref#1{\global\advance\refno by1
 \immediate\write\refs{\noexpand\item{\the\refno.}#1\hfil\par}}
\def\ref#1{\aref{#1}\the\refno}
\def\refname#1{\xdef#1{\the\refno}}

\newcount\exno
\exno=0
\def\Ex{\global\advance\exno by1{\noindent\sl Example \the\exno:

\nobreak\par\nobreak}}

\parskip=6pt
%===============================================================================

% PAGE TITRE
\Title{\vbox{\baselineskip12pt
\hbox{LAVAL-PHY-98-28}}}
{\vbox {\centerline{Generating functions for tensor products.}}}

\centerline{L. B\'egin$^\natural$\foot{Work supported by NSERC
(Canada).}, C. Cummins$^{\sharp 2}$ and P. Mathieu$^\natural$\foot{Work supported
by NSERC (Canada) and FCAR (Qu\'ebec).} }
\smallskip\centerline{$^\natural$ \it D\'epartement de Physique,
Universit\'e Laval, Qu\'ebec, Canada G1K 7P4}
\smallskip\centerline{$^\sharp$ \it CICMA, Department of Mathematics and Statistics, Concordia University,}
\centerline{Montr\'eal, Qu\'ebec, Canada H3G 1M8}

\noindent

\noindent{\bf Abstract}: This is the first of two articles devoted to a
comprehensive exposition of the generating-function method for computing fusion
rules in affine Lie algebras.  The present paper is entirely devoted to
the study of the tensor-product (infinite-level) limit of fusions rules.
We consider thus in detail the problem of constructing tensor-product generating
functions in finite Lie algebras.  From the beginning, the problem is recast in
terms of the concept of a {\it model}, which is an algebra whose Poincar\'e
series is the generating function under study.  We start by reviewing Sharp's
character method. Simple examples are worked out in detail,
illustrating thereby its intrinsic limitations.  An
alternative approach to the construction of tensor-product generating function is
then presented which overcomes most of the technical difficulties associated to
the character method. It is based on the reformulation of the problem of
calculating tensor products in terms of the solution of a set of linear and
homogeneous Diophantine equations
whose elementary solutions represent ``elementary couplings''.  Grobner bases
provide  a tool for generating the complete set of relations between elementary
couplings and, most importantly, as an algorithm for specifying a  complete,
compatible  set
of ``forbidden couplings''. 
This machinery is then applied to the construction of
various tensor-product generating functions.

\Date{11/98\ \ (hepth@xxx/9811113)}

\let\n\noindent

%============================================================================== 

\newsec{Introduction}

\subsec{Orientation}

Fusion rules yield the
number of independent couplings between three given primary fields in conformal
field theories.  We are interested in fusion rules in unitary  conformal field
theories that have a Lie group symmetry, that is, those whose generating
spectrum algebra is an affine Lie algebra at integer level. These are the
Wess-Zumino-Witten models [\ref{V.G. Knizhnik and A.B. Zamolodchikov, Nucl. Phys.
{\bf B247} (1984) 83.},\ref{D. Gepner and E. Witten, Nucl. Phys. {\bf B278} (1986)
493.}\refname\GW]. Primary fields in these cases are in 1-1 correspondence with
the integrable representations of the appropriate affine Lie algebra at level
$k$.  Denote this set by $P_+^{(k)}$ and a primary field by the corresponding
affine weight $\lah$.  Fusion coefficients ${\Nc_{\lah\muh}^{(k)}}~^{\nuh}$ are
defined by the product
$$\lah\times \muh = \sum_{\nu\in P_+^{(k)}} {\Nc_{\lah\muh}^{(k)}}~^{\nuh}
\; \nuh\eq$$

In the infinite-level limit and for fields with finite conformal dimensions (i.e.
finite values of the finite (non-affine) Dynkin labels -- see below), the purely
affine condition on weight integrability is relaxed and the primary fields are 
solely characterized by their finite part, required to be an integrable weight of
the corresponding finite Lie algebra. Recall that a weight ${\lambda}$  is
characterized  by its expansion coefficients in terms of the fundamental weights
${\omega}_{i}$
$${\lambda}=\sum_{i=1}^{r} \lambda_i
{\omega}_{i} = (\lambda_1,...,\lambda_r)\eq$$ where $r$ is the rank of the
algebra.  The numbers
$\la_i$'s are the Dynkin labels.  The set of weights with nonnegative Dynkin
labels (the integrable weights) is denoted by
$P_+$.  

Similarly, in the infinite-level limit, the fusion coefficients reduce to
tensor-product coefficients:
$$\lim_{k \rightarrow \y} {\Nc_{\lah\muh}^{(k)}}~^{\nuh}= 
{\Nc_{\lambda\mu}}^{\nu}.
\eqlabel\limformula$$
where ${\Nc_{\lambda\mu}}^{\nu}$ is defined by
$$\la\otimes \mu = \sum_{\nu\in P_+} {\Nc_{\lambda\mu}}^{\nu}\;  \nu\eq$$
By abuse of notation, we use the same symbol for the highest weight and
the  highest-weight representation. Notice that
$$\Nc_{\lambda\mu}~^{\nu}=\Nc_{\lambda\mu \nu^*}\eq$$
where $\nu^*$
 denotes the highest weight of the representation
conjugate to that of $\nu$. Equivalently, $\Nc_{\lambda\mu \nu^*}$ gives the
multiplicity of the scalar representation in the triple product
$\la\otimes\mu\otimes\nu^*$.

This paper is devoted to a detailed analysis of tensor products from
the point of view of generating-function methods, preparing
the ground for  an analysis of 
fusion rules along the same lines. This will be the
subject of a following article.

%--------------------------------------------------------------------------
\subsec{Tensor-product generating functions}

A tensor-product generating function codes the information for all the
tensor products of a given algebra in a single function.  This function is
first defined in terms of an infinite series as follows.  We introduce the
dummy variables $L_i, M_i, N_i$ with $i=1, \cdots, r$ and multiply the
tensor product coefficient ${\Nc_{\lambda\mu}}^{\nu}$ by $L^\la M^\mu N^\nu$ where
$$L^\la=L_1^{\la_1}\cdots L_r^{\la_r}\eq$$
and sum over all integrable values of $\la,\mu,\nu$:
$$G(L,M,N)= \sum_{\la,\mu,\nu \in P_+} {\Nc_{\lambda\mu}}^{\nu} L^\la M^\mu
N^\nu\eq$$
$G$ can generally be expressed in terms of a rather simple closed function of its
variables.  For instance, for $su(2)$, it reads
$$G(L,M,N)= {1\over (1-LM)(1-LN)(1-MN)}\eq$$

The above generating function  contains all the global information concerning
the tensor products in
$su(2)$. 
An example of  global information that can be extracted from such an
expression is the number of couplings having a
particular property. Suppose that we are interested in the total number of triple
products with $\la_1+\mu_1+\nu_1= 2n$ for a given positive integer value of $n$.
We can rescale the three dummy variables by a factor $s$, giving
$$\eqalign{ G(L,M,N; s) &= {1\over (1-s^2LM)(1-s^2LN)(1-s^2MN)}\cr & =
\sum_{n=0}^\infty s^{2n}\sum_{\la_1,\mu_1,\nu_1\geq 0}
{\Nc_{\lambda_1\mu_1}}^{\nu_1}
 L^{\la_1}M^{\mu_1}N^{\nu_1}\cr}\eq$$ We then set $L=M=N=1$ and read
the desired information from the coefficient of $s^{2n}$ in the expansion:
$$G(s)= {1\over (1-s^2)^3} =
\sum_{n=0}^\infty s^{2n}\sum_{\la_1+\mu_1+\nu_1=2n} {\Nc_{\lambda_1\mu_1}}^{\nu_1}
\eq$$
that is
$$\sum_{\la_1+\mu_1+\nu_1=2n} {\Nc_{\lambda_1\mu_1}}^{\nu_1} = {(n+2)(n+1)\over
2}\eq$$

Furthermore, the generating function contains local information. A simple example
of useful local information is the value of an explicit tensor-product
coefficient.
 Given a closed expression for the generating function,
explicit tensor-product coefficients can be read from its Taylor series
expansion. 

This in turn shows that another example of basic global information that
can be deduced from  a generating function is the integrality as well as the
positivity of the tensor-product coefficients.  In fact, tensor-product
generating functions
$G$
 can always be written in a manifestly positive form.

More
 importantly, from our point of view, is that  in the context of fusion rules, the
construction of the simplest generating functions led to the discovery of the
notion of threshold levels [\ref{C.J. Cummins, P. Mathieu and M.A. Walton, Phys.
Lett. {\bf B254} (1991) 390.}\refname\CMW].  Moreover, as  shown in the sequel
paper, setting up a fusion generating function is a way to obtain explicit
expressions for these threshold levels.  Our new approach to fusion-rule
generating functions, which originates from the generalization of techniques
developed in the present paper on tensor products, leads to a further new
concept, that of fusion basis.

%---------------------------------------------------------------------------
\subsec{Overview of the paper}

 The present article is organized as follows.  We start by explaining
  in detail the construction of tensor-product generating
functions for finite Lie algebras. The first construction  which
is presented is the character method developed by Sharp and his collaborators 
(section 2).  Here the starting point is the construction of the generating
function for the characters of all the irreducible representations which serves
as the building block for the construction of the tensor-product generating
functions. Simple examples of generating functions are worked out in details
($su(2),\, su(3),\, sp(4)$ and
$osp(1,2)$). We also introduce and exemplify the concept of   
a {\it model} for a generating function. This is an algebra
whose Poincar\'e series is the
generating function  under study. 
Models allow us to make use of powerful techniques from 
computational algebra to find generating functions as described below.

Although it is conceptually very simple,  the
character method is strongly limited by its inherent computational difficulties:
the disproportion between the simplicity of the resulting form of the generating
function and the intermediate calculations is enormous. This motivates our
alternative approach to the construction of
tensor-product generating function. 
It is based on the reformulation of the problem of calculating
tensor products in terms of the solution of a set of linear and homogeneous
Diophantine equations (cf. section 3).  The Diophantine-equation
reformulation of the problem is equivalent to its expression in terms of 
 counting
 the number of integer solutions in a convex polytope and the polytope equations
appropriate to any classical Lie algebra have been described by
Berenstein and Zelevinsky. For $su(N)$, these inequalities are equivalent to
the
Littlewood-Richardson algorithm, presented in section 4.   The elementary
solutions of these Diophantine equations represent  ``elementary couplings''. 
For $sp(4)$, the use of the
Berenstein-Zelevinsky inequalities to obtain the elementary couplings and their
relations (cf. the analysis of section 6) is new.

 The essential difficulty in constructing a generating function,
given a set of inequalities is not finding the elementary solutions.
For examples of the size we consider efficient algorithms exist
to solve this problem.
The key difficulty is related to
the numerous relations -- just alluded to -- that exist in general between the
elementary solutions. 
{}From the Diophantine-equation point of view, the decomposition of a solution may
not be unique because different sums of elementary solutions could yield the same
result.

These
redundancies must then be eliminated by forbidding one of the two sums of
elementary solutions 
occurring in each
relation.  However, the procedure for doing this when there are more that one
relation is somewhat tricky: we cannot select at random a sum  to
be forbidden from each relation and be sure that no further relations will
arise.  This is the problem of finding a compatible set of forbidden couplings. 
To solve this problem
we first ``exponentiate'' it: given a solution $\alpha =
(\alpha_1,\dots,\alpha_k)$ to our system of linear Diophantine
equations, we introduce formal variables $X_1,\dots,X_k$
and consider the monomial $X_1^{\alpha_1} \dots X_k^{\alpha_k}$.
The linear span, $R$,
of all such monomials is a model for the
generating function for the solutions to the original
set of linear Diophantine equations (see section 5).
To find the Poincar\'e series of $R$ we use
the theory of Grobner bases.  Given a set of elementary
couplings, or in this case generating monomials, 
Grobner basis techniques provide  a  method for
generating a complete set of relations. In addition, and more importantly, they
provide an algorithm for specifying a compatible set of ``forbidden couplings''. 
In other words one can calculate the Poincar\'e series of $R$ and
hence the required generating function, directly from the set of
relations provided by the Grobner basis method.
It is then shown how this machinery  applies to the construction of
tensor-product generating functions. This is the subject of section 5.

  A trivial modification of the description of tensor products in terms of
inequalities follows by introducing extra positive labels that transforms the
inequalities into equalities.  For $su(N)$, this gives rise to a remarkable
graphical construction, the famous
Berenstein-Zelevinsky triangles.  These are introduced in section 7.  We also
discuss the analogous construction for $sp(4)$, whose diagrammatic representation
is new. But the main interest of these reformulations is that it yields a
simple and systematic way of obtaining the elementary couplings from the
construction of a vector basis.  A  projection operation yields the
elementary couplings. In the process, we get a new way of constructing the
generating functions. This  vector-basis approach to the construction of
tensor-product generating functions is illustrated with $su(3)$ and $sp(4)$.

%------------------------------------------------------------------

\subsec{What is new?}

In part, this article is a comprehensive review of existing techniques for
constructing tensor-product generating functions.
 In addition, the
efficiency and the power of the different methods of construction are compared. 
We also, from a practical point of view, discuss the
connections between 
commutative algebra and the computation
and properties of generating functions. On the mathematical
side, these links have been developed  by Stanley 
[\ref{R.P. Stanley, Duke Math. J. {\bf 40} (1973) 607; 
{\it Combinatorics and Commutative Algebra}, (Boston:
Birkhauser) (1983).}\refname\Stan]. 
In this article, emphasis is  placed on the  concept of a {\it
model} of a generating function, which
is an algebra whose Poincar\'e series is equal to the given generating
function.   This  allows us to make use of 
powerful techniques in computational commutative algebra including
Grobner bases,  which were first used in this context in 
[\ref{M.Couture,
C.J.Cummins and R.T.Sharp, J.Phys {\bf A23} (1990) 1929.}\refname\CCS].

In this way we arrive at new derivations of previously known results.  But those
pertaining to $sp(4)$, and in particular the use of the Berenstein-Zelevinsky
inequalities as a way to fix unambiguously the linear relations between the
``elementary couplings'', are new and should be stressed.  However, the main
achievement is a sound reformulation of the problem that is well-adapted to its
extension to the fusion case.

%============================================================================== 

\newsec{Generating-function for tensor products: the character method}

%------------------------------------------------------------------

\subsec{The character method for the construction of the tensor-product
generating function: the 
$su(2)$ case}

The method developed by Sharp and collaborators for constructing generating
functions for tensor products is based on manipulations of the character
generating functions [\ref{R.
Gaskell, A. Peccia and R.T. Sharp, J. Math. Phys. {\bf 19} (1978)
727; J. Patera and R.T. Sharp, in {\it Recent advances in group theory and
their applications to spectroscopy}, ed. J. Domini, New-York, Plenum; J. Patera
and R.T. Sharp, in Lecture Notes in Physics (New York, Springer Verlag, 1979)
vol. 94, p. 175;  M.
Couture and R.T. Sharp, J. Phys. {\bf A13} (1980) 1925; R.T. Gaskell and R.T.
Sharp, J. Math. Phys. {\bf 22} (1981) 2736;  C. Bodine and R.T. Gaskell, J. Math.
Phys. {\bf 23} (1982) 2217;   R.V. Moody, 
J. Patera and R.T. Sharp, J. Math. Phys. {\bf 24} (1983) 2387; 
J. Patera and R.T. Sharp, J. Phys. {\bf A13} (1983) 397;  Y. Giroux, M. Couture and R..T. Sharp, J. Phys. {\bf
A17} (1984) 715.}\refname\SP]. Although
simple in principle, these manipulations become rather cumbersome as the rank of
the algebra is increased. To illustrate the method, we will work in complete
detail the simplest example, the
$su(2)$ case.

The first step is the derivation of the character generating function.
The
Weyl character formula for a general algebra of rank $r$ and a  highest-weight
representation
$\lambda$ is
$$\chi_\lambda= {\xi_{\lambda+\rho}\over \xi_\rho}\eq$$ where $\rho$ is the
finite Weyl vector, $\rho= \sum_{i=1}^r\om_i$,
and where the characteristic function $\xi$ is defined as
$$\xi_{\la+\rho}=\sum_{w\in W}\, \epsilon(w)e^{w(\la+\rho)}\eq$$
where $\epsilon(w)$ is the signature of the Weyl reflection $w$ and $W$ stands for
the Weyl group.  

For $su(2)$, the Weyl group contains two elements: $1,s_1$.  With $$x=
e^{\om_1}\eq$$ the 
$su(2)$ characteristic function $\xi$ for the representation of highest weight
$m\om_1\equiv(m)$ is
$$\xi_{m+1} = x^{m+1}-x^{-m-1}\eq$$
The character reads then
%àà add an intermediate equality
$$\chi_m = { x^{m+1}-x^{-m-1}\over  x-x^{-1}}={ x^{m}-x^{-m-2}\over 
1-x^{-2}}=  x^m+x^{m-2}+\cdots+ x^{-m}\eqlabel\aa$$ The character generating
function
$\chi_L$ is obtained by multiplying the above expression by $L^m$ where $L$ is a
dummy variable, and summing over all positive values of
$m$: 
$$\eqalign{ \chi_L(x) &= \sum_0^\y L^m\chi_m = {1\over x-x^{-1}} \sum_0^\y
L^m(x^{m+1}-x^{-m-1})\cr &= {1\over 1-x^{-2}}\left({1\over 1-Lx}-{x^{-2}\over
1-Lx^{-1}}\right) = {1\over (1-Lx) (1-Lx^{-1})}\cr}\eq$$
By construction, the character of the highest weight $(m)$ can be
recovered from the power expansion of $\chi_L$ as the coefficient of the term
$L^m$. The characteristic generating function $\xi_L$ is defined by
$$\chi_L(x) = {\xi_L\over \xi_0}\eq$$
and it reads
$$ \xi_L(x) = {x-x^{-1}\over (1-Lx) (1-Lx^{-1})}={x\over 1-Lx}-{x^{-1}\over
1-Lx^{-1}}\eq$$
the last form being the one that results directly  from (\aa).  

The tensor product of two highest-weight representations can be obtained from
the product of the corresponding characters:
%* add the factor {\Nc_{mn}}^\ell:
$$\chi_m \chi_n= \sum_\ell  {\Nc_{mn}}^\ell\;\chi_\ell\eq$$  
This information can be extracted
from the product of the corresponding generating functions.  We are thus led to
consider the product $\chi_L (x)\chi_M(x)$.  To simplify the analysis of the
resulting expression, notice that the information concerning the
representations occurring in the tensor product is coded in the leading term of
the character, i.e., the term $x^{m+1}$.  The rest of the representation  is
easily reconstructed by the application of the Weyl group and the action of the
ladder operator.  Actually, to insure that every positive
power of $x$ singles out a highest-weight representation, we can multiply both sides by
$\xi_0$:   To read off these terms, we can focus on the terms
with strictly positive powers of
$x$ in the product 
$\chi_L(x)
\chi_M (x)\xi_0(x)$.
If we want to read off directly the Dynkin label of the representations (and
not their shifted value), it is  more convenient to divide  by $x$ before doing
the projection, now restricted to the nonnegative powers of $x$.  The
truncation of an expression by its negative powers of $x$ will be denoted by the
MacMahon symbol [\ref{P. MacMahon, {\it Combinatory analysis}, 2 vols
(1917,1918), reprinted by Chelsea, third edition, 1984.}\refname\Mac]
$\Omega$, defined by
$$\Ox \,\sum_{-\y}^\y c_nx^n = \sum_{n\geq0} c_nx^n\eq$$
When there is no ambiguity concerning the variable in terms of which the
projection is defined, it is omitted from the $\Omega$ symbol.

We are thus interested in the projection of the following expression
$$\eqalign{ \chi_L(x) \chi_M(x) \xi_0 (x)\x &=  \chi_L(x)
\xi_M(x) \x \cr &= {1\over (1-Lx) (1-Lx^{-1})}\left({1\over
1-Mx}-{x^{-2}\over 1-Mx^{-1}}\right)\cr}
\eqlabel\twopiece$$ For these manipulations, we use systematically the following
simple identities:
$$\eqalign{ {1\over (1-Ax) (1-Bx^{-1})} &= {1\over (1-AB)} \left({1\over
1-Ax} + {Bx^{-1}\over 1-Bx^{-1}}\right)\cr &=  {1\over (1-AB)} \left({Ax\over
1-Ax} + {1\over 1-Bx^{-1}}\right)\cr
&=  {1\over (1-AB)} \left({1\over
1-Ax} + {1\over 1-Bx^{-1}} -1\right)\cr}\eq$$
%àà add an equality above and a comment below
Which one is used is dictated by the context (a good choice often simplifies the
subsequent analysis, the last one being somewhat more algorithmic).

We thus have two terms to analyze.  The first one is
$${1\over (1-Lx) (1-Lx^{-1})(1-Mx)}= 
{1\over (1-Lx) (1-LM) }\left( {1\over 1-Mx} +{Lx^{-1}\over 1-Lx^{-1}} \right)\eq$$
The first part is not affected by the projection. The second one can be
written as
$${Lx^{-1}\over (1-Lx) (1-LM)(1-Lx^{-1}) } = {Lx^{-1}\over
(1-LM)(1-L^2)}\left({Lx\over 1-Lx}+ {1\over 1-L\x}\right)\eq$$
The second part of the above expression contains only negative powers of $x$ and
can thus be ignored. The first part is unaffected by the projection. We have thus,
for the first term of (\twopiece)
$$\O\, {1\over (1-Lx) (1-Lx^{-1})(1-Mx)} =  {1\over (1-Lx) (1-LM) }\left( {1\over
1-Mx}+{L^2\over1-L^2}\right)\eqlabel\prem$$
The other term that needs to be considered is
$$\eqalign{& \O\, {x^{-2}\over (1-Lx) (1-Lx^{-1})(1-Mx^{-1}) }\cr
 \qquad\qquad \quad &= \O\, {x^{-2}\over 
(1-Lx^{-1})(1-LM)}\left( {1\over 1-Lx}+{M\x \over 1-Mx^{-1} }\right)\cr
\qquad \qquad \quad &= \O\, {x^{-2}\over 
(1-Lx^{-1})(1-LM) (1-Lx) }\cr
\qquad\qquad \quad &= \O\, {x^{-2}\over 
(1-LM)(1-L^2) }\left( {Lx\over 1-Lx}+{1\over 1-L\x}\right)\cr
\qquad\qquad \quad &= \O\, {Lx^{-1}\over 
(1-LM)(1-L^2)(1-Lx)}\cr
\qquad \qquad \quad &= {Lx^{-1}\over 
(1-LM)(1-L^2)}\left( {1\over 1-Lx} -1\right) \cr
 \qquad\qquad \quad &=  {L^2\over 
(1-LM)(1-L^2)(1-Lx)}\cr}  \eqlabel\seco$$
Subtracting (\seco) from (\prem), we find that 
$$\O\, \chi_L(x) \xi_M(x)\,\x\, = {1\over 
(1-LM)(1-Lx)(1-Mx)}\eq$$
Replacing $x$ by $N$, we thus get
$$G^{su(2)}(L,M,N)= {1\over 
(1-LM)(1-LN)(1-MN)}\eqlabel\sufun$$
as the generating function for $su(2)$
 tensor products.  To read off the
representations $(n)$ that occur in the tensor product $(\ell)\otimes(m)$, we
expand the generating function (as a Taylor series) and collect all terms $N^n$ 
that are multiplied by $L^\ell M^m$.  All such values of $n$ are the
representations we are looking for. This is simply a restatement of the
following relation between the tensor-product coefficients ${\Nc_{\ell m}}^n$
and the generating function:
$$G(L,M,N)= \sum_{\ell, m,n \geq 0}{\Nc_{\ell m}}^n L^\ell M^mN^n\eq$$

%-----------------------------------------------------------------

\subsec{The abstract setting: Poincar\'e series, elementary couplings and
relations; defining a model}

 As we shall see it is frequently useful
have a {\it model},
$R$,  for a generating function $G(X_1,\dots,X_k)$ such as (\sufun). 
By this we mean
a commutative $\Q$-algebra with an identity, graded by $\N^k$, 
$$R=\oplus_{\alpha\in \N^k} R_\alpha\, , \qquad \quad R_\alpha R_\beta\subseteq
R_{\alpha+\beta}\eq$$  
and such that its  Poincar\'e series
\foot{Such series are also called Hilbert series.}
$$
F(R) = \sum_{\alpha\in \N^k} \dim_\Q(R_\alpha) X^\alpha 
$$
satisfies
$$
F(R)  = G(X_1,\dots,X_k).\eq
$$
For example, for (\sufun), with $X_1= L, \, X_2= M, \, X_3=N$,  we can take
$R=\Q[E_1,E_2,E_3]$  (in fact
all our examples $R$ is either a subring or quotient of a polynomial) with
the grading of $E_1,E_2$ and $E_3$ being
$(1,1,0)$, $(1,0,1)$ and $(0,1,1)$. The homogeneous subspaces
are spanned by $E_1^a E_2^b E_3^c$, $a,b,c\in \N$ with grade
$(a+b,a+c,b+c)$ and so
$$
F(R) = \sum_{(a,b,c)\in\N^3} X_1^{a+b}X_2^{a+c}X_3^{b+c}
= G^{su(2)}(X_1,X_2,X_3)
\eq$$
as required.

If $R$ is generated by elements $E_1,\dots,E_s$
and is a model for a generating function $G$ for tensor products
(or fusion products)
 then we
call $E_1,\dots,E_s$ a set of ``elementary couplings''
for $G$.

It should perhaps be stressed that {\it a priori}
the  variables 
$X_1,\dots,X_k$ and
$E_1,\dots,E_s$ are unrelated. 
We shall refer to the $E$'s as {\it model variables}
and the $X$'s as {\it grading variables}.
If the grading vector of $E_i$ is $\alpha^i$ , $i=1,\dots,s$
then 
there is an associated monomial in 
the grading variables: $X^{\alpha^i}$, for which we
will use the notation $g(E_i)$.
For example in 
the above example we have $g(E_1) = X_1^1X_2^1X_3^0 = LM$.
However, to avoid tedious repetition when writing down
generating functions we shall often write, for example,
$1/(1-E_1)$ rather than $1/(1-g(E_1))$. In {\it all} such
cases where model variables appear in a generating function
they should be replaced by the corresponding monomial
in the grading variables. The utility of this convention
can be seen by examining some of the examples of generating
functions contained in the following sections and comparing
them with the much more unwieldy expressions obtained by
everywhere replacing the model variables by grading variables.

In the  case of tensor products we use
the notation ``$E : g(E) : \hbox{\rm product}$''
to denote a set of elementary couplings with their ``exponentiated'' grading 
and the corresponding term in 
 the tensor product. So in the example above we would write:
 $$\eqalignD{ &E_1: LM: \quad &(1)\otimes(1)\supset(0), \cr & E_2:
 LN: &(1)\otimes(0)\supset(1),\cr & E_3: MN: &(0)\otimes(1)\supset(1)\cr}
 \eqlabel\elesud$$ 

Having made the distinction  between
grading and model variables, it should be noted that 
there are 
cases where we can identify the model as a ring 
generated by monomials in the grading variables.
So in the above example we could {\it define}
$E_1=LM$, $E_2=LN$ and $E_3=MN$ and take the
model for our generating function to be the
subring of $Q[L,M,N]$ generated by $E_1,E_2$ and
$E_3$. However, it is not always desirable, or
even possible, to make this identification. 

As we shall see,
identifying a generating function with the Poincar\'e series
of a model  will allow us to link the generating function
more explicitly with combinatorial rules for calculating
tensor product (and other) multiplicities. It will also allow
us to make use of powerful algebraic techniques for
computing Poincar\'e series.

In the  example above we are given a generating function from which it
is easy to find a model. Of course in general we will start with a model
from which we construct the required generating function. We close
this
section with two examples of how models for the $su(2)$ character
generating function can be constructed.

The first method, which has been exploited by Sharp et al (see
[\SP]) to
construct character generating functions, amounts to finding an algebra $R$ which
is a module for the Lie algebra $su(2)$ and such that as
an $su(2)$ module $R$ is isomorphic to $\oplus_{i\geq 1} V_i$ where
$V_i$ is the irreducible $su(2)$ module of dimension $i$.

In this case we can take $R=\Q[p,q]$ with the generators of
$su(2)$ being given by differential operators: 
$$
h=p{{\partial}\over{\partial p}} -q{{\partial}\over{\partial q}},\quad
x_- = q{{\partial}\over{\partial p}},\quad
x_+ = p{{\partial}\over{\partial q}}\eq
$$
The $su(2)$ highest-weight vectors are $p^i$, $i\geq 0$ and a basis
of the irreducible submodule of dimension $i$ is just given by
the monomials of degree $i$ in $p$ and $q$. We can give $R$ an  $\N^3$ grading
by taking the degree of $p$ to be $(1,1,0)$ and of $q$ to be $(1,0,1)$.
%+ I would like to add a word to explain the 3 indices:
Here the first grading index specifies the representation while the other two
refer to a particular weight. As $R=\Q[p,q]$ the Poincar\'e function for $R$
is, 
$${1\over {(1-p)(1-q)}}$$
with the understanding, as explained above,
that $p$ and $q$ should be replaced
by the corresponding expression in terms of the grading variables.
Let us denote these grading variables here by $L$ (which exponentiates the
representation index) and $x,y$ (exponentially related to the weights).  The
Poincar\'e function reads then
$${1\over {(1-Lx)(1-Ly)}}\eq$$
Of course in this case the distinction involved and the
usefulness of our convention is not immediately apparent.

Note that in this example we could take a
$\Z^2$ grading with the grades of $p$ and $q$ being $(1,1)$ and $(1,-1)$. However
in general allowing negative weights means we have to be much
more careful when taking Laurent series expansions and so we
avoid this complication when possible.   This grading is related to a single
variable description of the weights $x^i, \cdots , x^{-i}$. Setting $y=x^{-1}$
 gives the corresponding form of the generating function.

Another way of constructing a model for the weight generating function, 
 which
makes more natural the $\N^3$ grading,  is to observe that the complete set
$SU(2)$ weight vectors  of
finite dimensional irreducible $su(2)$  modules 
are in 1-1 correspondence
with one-rowed Young tableaux. If the Young  tableau has $c$ boxes
filled with $a$ 1's and $b$ 2's then there is a constraint
$$a+b-c=0,\quad a,b,c\geq 0\eqlabel\lind$$
and so the solutions to this linear
Diophantine equation are in 1-1 correspondence with the complete
set of  $SU(2)$ weight vectors. Thus to find a model for the
weight generating function it is sufficient to find a model
for the solutions to (\lind). In this example
it is not difficult to see that every solution to this equation
is a linear combination (with non-negative coefficients)
of the two fundamental solutions:
$(a,b,c) = (1,0,1)$ and $(a,b,c) = (0,1,1)$. Let $R$ be the
subring of $\Q[A,B,C]$ generated by the monomials $E_1=AC,\, E_2=BC$.
Considering the exponents of the monomials $E_1$ and $E_2$, we
see that the monomials in $R$ correspond to the solutions of 
(\lind) and hence taking the 
natural grading on $R$
ensures that the Poincar\'e series of $R$ is the generating
function for the solutions to  (\lind) and hence is the required
generating function. In this example there are no relations between
$E_1$ and $E_2$ and so $R$ is isomorphic to the polynomial ring 
in
two variables (as expected) and so the Poincar\'e function is once
again
(with $A\rw x, B\rw y, C\rw L$):
$${1\over {(1-Lx)(1-Ly)}}\eq$$

In all these simple examples we have found a model which is
a polynomial ring. In general this will not be the case. This
is illustrated in the next section.
%------------------------------------------------------------------

\subsec{Multiple $su(2)$ tensor products}

In order to illustrate in a rather simple context the occurrence of 
relations between elementary couplings, we will consider a slight
generalisation of the previous problem namely,  finding the multiplicity of a
given representation $\zeta$ in the triple product
$\la\otimes \mu\otimes \nu$. In terms of character generating functions, this
amounts to considering the product
$\chi_L(x)\chi_M(x)\chi_N(x) \supset \chi_P(x)$,
or equivalently, $\chi_L(x)\chi_M(x)\xi_N(x)\x \supset \xi_P(x)\x$. The left
side is
then projected onto positive powers of $x$. We are thus led to consider
 $$\O\; {1\over (1-Lx)(1-L\x)(1-Mx)(1-M\x)}\left({1\over 1-Nx}- {x^{-2}\over
1-N\x}\right)\eq$$
The projection of each term is worked out as previously and the
resulting expression is found to be, with $x$ replaced by $P$:
$$G(L,M,N,P) = {1-LMNP\over (1-LP)(1-MP)(1-NP)(1-LM)(1-LN)(1-MN)}\eqlabel\ffa$$
This is the sought for generating function. Here we would like
to have a model with 6 elementary couplings corresponding
to the terms in the 
denominator of the generating function:
$$\eqalignD{
&E_1: LM: \quad &(1)\otimes(1)\otimes(0)\supset(0)\cr
&E_2: LN:  &(1)\otimes(0)\otimes(1)\supset(0)\cr
&E_3: LP:  &(1)\otimes(0)\otimes(0)\supset(1)\cr
&E_4: MN:  &(0)\otimes(1)\otimes(1)\supset(0)\cr
&E_5: MP:  &(0)\otimes(1)\otimes(0)\supset(1)\cr
&E_6: NP:  &(0)\otimes(0)\otimes(1)\supset(1)\cr}\eq$$
and there must be a linear relation between the following products 
(signalled by a term in the numerator)  
which have grading $LMNP$:
\foot{Relations between products of
elementary couplings are often called {\it syzygies} in the physics literature (see
in particular [\SP] and related works).  However, the proper mathematical meaning of
a syzygy is somewhat different and for this reason we stick to the more correct
terminology of {\it relation}.}
$$E_1E_6, \qquad E_2E_5, \qquad E_3E_4\eq$$

It is not difficult to see that a model is given by
$\Q[e_1,e_2,e_3,e_4,e_5,e_6]/I$ where $E_i=e_i+I, i=1,\dots,6$
and $I=(ae_1e_6 + be_2e_5 +c
e_3e_4)$ is the ideal generated by the polynomial $ae_1e_6 +
be_2e_5 +ce_3e_4$ with $a,b,c\in\Q$ not all zero. In particular
the model is far from unique. As we shall see later, particular
methods of construction will select one particular model.

Before leaving this example, we would like to rework it from a different
point of view, as an illustration of the `composition' technique of generating
functions.
 Let $G(L,M,R)$ describe the tensor product corresponding to
$\chi_L\chi_M\supset\chi_R$ and similarly let $G(Q,N,P)$ correspond to
$\chi_Q\chi_N\supset\chi_P$. We are interested  the product
$\chi_L(x)\chi_M(x)\chi_N(x) \supset \chi_P(x)$ but treated from the product
of the two generating functions $G$.  We thus want to enforce the constraint $R=Q$
in the product 
$G(L,M,R)G(Q,N,P)$. 
The idea\foot{This trick is used in different references in [\SP], mainly in
relation with the construction of generating functions for branching functions.}
is to multiply this product by
$(1-Q^{-1}R^{-1})^{-1}$ and, in the expansion in powers of $R$ and $Q$, keep only
terms of order zero in both variables: with an obvious notation we have
$$ \eqalign{\OR\OQ\, & G(L,M,R)G(Q,N,P) {1\over1-Q^{-1}R^{-1}}\cr
  &\qquad\,=\, \OR\OQ\,\sum_n A_n(L,M)\, R^n \sum_m B_m(N,P)\,Q^m\sum_\ell
R^{-\ell}Q^{-\ell}\cr&\qquad\,= \sum_p A_p(L,M) B_p(L,M)\cr}\eq$$
which is manifestly equivalent to considering
$$ \Oxx\,G(L,M,x)G(\x,N,P) \eq$$
With the explicit expressions for the generating functions, we have thus
$$\Oxx\, {1\over (1-Lx)(1-Mx)(1-LM)} {1\over (1-P\x)(1-N\x)(1-NP)}\eq$$ A brief
 and by now standard analysis yields directly the generating function (\ffa). 

%------------------------------------------------------------------

\subsec{The $osp(1,2)$ case}

The simplest example after $su(2)$ is that of the superalgebra $osp(1,2)$. Very
little information is needed about superalgebras for the study of the
 $osp(1,2)$ representations.  We only need the fact that the highest
weight $osp(1,2)$ representations $\{m\}$ decompose into a direct sum of two
$su(2)$ representations; $(m)\,\oplus\, (m-1/2)$ [\ref{M. Scheunert, W. Nahm and
V. Rittenberg, J. Math. Phys. {\bf 18} (1977) 155; M. Marcu, J. Math. Phys. {\bf
21} (1980) 1277.}].  The character is thus
$$\chi_{\{m\}}= {x^{m+1}+x^m-x^{-m}-x^{-m-1}\over x-\x}\eq$$
It is a simple exercise to check that the generating character function is
$$\chi_L= {1+L\over (1-Lx)(1-L\x)}\eq$$
The tensor-product generating function is found to be [\ref{R.T. Sharp, J. Van der
Jeught and J.W.B. Hugues, J. Math. Phys. {\bf 26} (1985) 901.}]
$$\eqalign{ G^{osp(1,2)}(L,M,N)& = {1-(LMN)^2\over
(1-LM)(1-LN)(1-MN)(1-LMN)}\cr &={1+LMN\over (1-LM)(1-LN)(1-MN)}\cr} \eq$$ 
An underlying model would thus have four elementary couplings:
$$\eqalignD{ & E_1: LM: &\{1\}\otimes\{1\}\supset\{0\}\cr & E_2:
LN: &\{1\}\otimes\{0\}\supset\{1\}\cr & E_3: MN: &\{0\}\otimes\{1\}\supset\{1\}\cr
& E_4: LMN:&\{1\}\otimes\{1\}\supset\{1\} \cr} \eq$$
and the numerator indicates a linear relation between products of elementary
couplings of degree $(LMN)^2$ namely:
$$E_1E_2E_3\quad\hbox{\rm and}\quad E_4^2\eq$$

A model is given by
$\Q[e_1,e_2,e_3,e_4]/I$ where $E_i=e_i+I, i=1,\dots,4$
and $I=(ae_1e_2e_3 + be_4^2) $ is the ideal generated by the polynomial
$ae_1e_2e_3 + be_4^2$ with $a,b\in\Q$, but not both zero.

%----------------------------------------------------------------------------------
\subsec{The $su(3)$ case}

 The next example in complexity is $su(3)$.   
With
$$x_i= e^{\om_i}\qquad i=1,2\eqlabel\defxi$$ the characteristic function for a
representation of highest weight $(m,n)=m\om_1+n\om_2$ is
$$\eqalign { \xi_{(m,n)}
&=x_1^{m+1}x_2^{n+1}-x_1^{-m-1}x_2^{m+n+2}-x_1^{n+m+2}x_2^{-n-1}\cr
\qquad &
+x_1^{n+1}x_2^{-m-n-2}+x_1^{-m-n-2}x_2^{m+1}-x_1^{-n-1}x_2^{-m-1}\cr}\eq$$The
characteristic generating function is obtained by multiplying this result by
$L_1^mL_2^n$ and summing over all positives values of $m,n$. The result is
$$\xi_{L_1,L_2}= {1-L_1L_2\over (1-L_1x_1) (1-L_1x^{-1}_1x_2) (1-L_1x^{-1}_2)
(1-L_2x_2) (1-L_2x_1x^{-1}_2) (1-L_2x^{-1}_1)}\eq$$

The construction of the tensor-product generating function proceeds exactly as
for $su(2)$, but here it is much more complicated from the simple fact that
there are two variables. The result is [\SP]
$$\eqalign{ 
&G^{su(3)}(L_1,L_2,M_1,M_2,N_1,N_2)=(1-L_1L_2M_1M_2N_1N_2)\cr
&\times [(1-L_1 N_2)(1-L_1M_2)  (1-L_2 M_1) (1-L_2 N_1) ]^{-1}\cr&\times[(1-M_2
N_1) (1-M_1 N_2)(1-L_1 M_1 N_1) (1-L_2 M_2 N_2)]^{-1}\cr}\eq$$
 {}From the denominator we see that it is natural to 
seek a model with
eight elementary couplings:
$$\eqalignT{&E_1  : L_1 M_2 ,\qquad\quad & E_2 : L_1 N_2 ,\qquad\quad 
&E_3 : M_1 N_2\cr
&E_4  : L_2 M_1 ,\quad &E_5 : L_2 N_1 ,\quad &E_6 : M_2 N_1 \cr
&E_7  : L_1 M_1 N_1,\quad  &E_8 : L_2 
M_2 N_2 & \cr} \eqlabel\eletrois$$ 
The numerator indicates a relation between these elementary
couplings that need to be taken into account to avoid over-counting multiplicities:
there must be a linear relation between the following three products:
$$E_1E_3E_5, \qquad E_2E_4E_6 \quad {\rm and}\quad E_7E_8\eq$$
which are the three terms with grading $L_1L_2M_1M_2N_1N_2$.
This is the only relation required and 
a model for this generating function is given by
$R = \Q[e_1,\dots,e_8]/I$, with $E_i=e_i+I, i=1,\dots,8$
and $I=(ae_1e_3e_5 + b e_2e_4e_6 + c e_7e_8)$
with $ae_1e_3e_5 + b e_2e_4e_6 + c e_7e_8\neq 0$.

The elements of $R$ have the form $m + I$ with $m\in
\Q[e_1,\dots,e_8]$. However there is no canonical way
of choosing the representatives $m$. Take for example
the case $a=b=c=1$. (Usually we will construct a model 
for our generating function as explained above and this
construction will fix the values of $a,b$ and $c$).
 In $R$ we have $ E_1E_3E_5 = -(E_2E_4E_6 + 
E_7E_8)$ and so we can take as a basis for $R$ the
set of (equivalences classes of ) monomials which do not contain the product $E_1E_3E_5$. In this case we say that we have chosen to make
$E_1E_3E_5$ a `forbidden product'.
Similarly we can forbid the products $E_2E_4E_6$ or
$E_7E_8$. As we shall see later, the choice of forbidden products
corresponds to a choice of {\it term ordering}.

We
can write the generating function in three different ways, each making manifest
the fact that one of the above products never appears. 
Note once again we use the convention that $E_i$ should be
replaced by $g(E_i)$ in these expressions:
$$\eqalign{ {G}&=\left(\prod_{i\not=1,3,5}~(1- E_i)^{-1}\right) \left(1+{ E_1 \over
(1- E_1) (1- E_5)} \right.\cr & \left.\qquad\qquad +{ E_3  \over (1- E_3)
(1- E_1)}+{ E_5
\over (1- E_5) (1- E_3)}\right)\cr
&=\left(\prod_{i\not=2,4,6}~(1- E_i)^{-1}\right) \left(1+{ E_2 \over
(1- E_2) (1- E_4)} \right.\cr & \left.\qquad\qquad+ { E_4  \over (1- E_4)
(1- E_6)}+{ E_6 \over (1- E_6) (1- E_2)}\right)\cr
 &=\left(\prod_{i\not=7,8}~(1- E_i)^{-1}\right) \left(1+{ E_7
\over(1- E_7)}+{E_8 \over (1- E_8)}\right) \cr}\eqlabel\fgsutr$$ 
It is clear 
that in expanding the first form, we
will never encounter a term corresponding to
a product of the three factors $E_1 E_3
E_5$. Similarly no product $E_2E_4E_6$ corresponds to a term in  the second form, while the
last expression amounts to forbidding all factors containing $E_7 E_8$.
Therefore, although
$G$ is unique, its expression in terms of the $ E_i$'s is not,
because the forbidden couplings may be chosen in different ways.

Before leaving this example, let us mention another way of
constructing the generating
function using the idea of `composition' of generating functions
described previously.
This uses the Giambelli formula that expresses a
general
representation in terms of a
difference of products of representations with a
single non-zero
Dynkin label,
i.e.,
$$(\mu_1,\mu_2) = (\mu_1+\mu_2,0)\otimes
(\mu_2,0)-
 (\mu_1+\mu_2+1,0)\otimes (\mu_2-1,0)\eqlabel\giam$$
%++ I have exchanged M_1 and M_2, since I think
%++ this way M_i carries the labels for \mu_i

First consider the generating function:
$$
G(L_1,L_2,M_1,N_1,N_2)\eq
$$
which is the generating function for products
of the form:
$(\la_1, \la_2)\otimes (\mu_1,0)$.
{}From this generating function we form:
$$
H(L_1,L_2,M_1,M_2,R_1,R_2)=
\ON\, 
G(L_1,L_2,M_1,N_1,N_2)G(N_1^{-1},N_2^{-1},M_2,R_1,R_2)\eq
$$
which is the generating function for
products of the form
$$(\la_1, \la_2)\otimes (\mu_1,0)\otimes
(\mu_2,0)\eq$$
Note that the generating function for
products $$(\la_1, \la_2)\otimes (\mu_1+1,0)\otimes
(\mu_2-1,0)\eq$$ is $ HM_2M_1^{-1}$ and so, by (\giam),
 the generating function for products
$(\la_1, \la_2)\otimes (\mu_1,\mu_2)$ is:
$$
\OM\, [ H - HM_2M_1^{-1}]\eq
$$
The coefficient of $M_1^{\mu_1}M_2^{\mu_2}$ is the
multiplicity of the representation with Dynkin
labels
$(\mu_1-\mu_2,\mu_2)$ in the product $$(\la_1,
\la_2)\otimes
\left[ (\mu_1,0)\otimes (\mu_2,0) -
 (\mu_1+1,0)\otimes (\mu_2-1,0)\right]\eq$$ To change to variables which carry
the Dynkin labels we make the substitution
$M_2 \mapsto M_2M_1^{-1}$, so that $M_1$ now
carries
the first Dynkin label. This introduces negative
powers of $M_1$, corresponding to products
$$(\la_1,
\la_2)\otimes 
\left[ (\mu_1,0)\otimes (\mu_2,0) -
 (\mu_1+1,0)\otimes (\mu_2-1,0)\right]\eq$$
 with $\mu_1 < \mu_2$,
which
are not required. So we must keep only positive
degree
terms in $M_1$ to obtain the final generating
function.

\subsec{The $sp(4)$ case}

As a final example, consider the $sp(4)$ case.  With the $x_i$ defined as in
 (\defxi),
the characteristic function is found to be 
$$\eqalign { \xi_{(m,n)}
&=x_1^{m+1}x_2^{n+1}-x_1^{-m-1}x_2^{m+n+2}-x_1^{n+m+5}x_2^{-n-1}
+x_1^{m+2n+3}x_2^{-m-n-2}\cr
 &+x_1^{-m-2n-3}x_2^{n+m+2}-x_1^{m+1}x_2^{-m-n-2}
-x_1^{-m-2n-3}x_2^{n}+x_1^{-m-1}x_2^{-n-1}\cr}\eq$$
and the characteristic generating function is
$$\eqalign{ \xi_{L_1,L_2}& = {1\over (1-L_1x_1) (1-L_1x_1x^{-1}_2) (1-L_2x^{-1}_2)
(1-L_2x_1^{-2}x_2)}\cr
&\times\left( {1+L_2\over (1-L_2x_1^2x^{-1}_2)(1-L_2x^{-1}_2)}+
{(1+L_2)L_1x_1\over (1-L_1x_1)(1-L_2x_1^2x^{-1}_2)} \right.\cr
& \left. \qquad\qquad
 +{L_1x^{-1}_1x_2\over
(1-L_1x_1)(1-L_1x^{-1}_1x_2)}\right)\cr}\eq$$
{}From this characteristic generating function, we construct the character
generating function and then we can proceed to the tensor-product generating
function.  This is again extremely cumbersome. The result is 
[\ref{M. Hongoh, R.T. Sharp and D.E. Tilley, J.Math. Phys. {\bf 15}
(1974) 782.}\refname\Hongo]
$$\eqalign{ &G^{sp(4)}(L_1,L_2,M_1,M_2,N_1,N_2)\cr &=[(1-M_1 N_1) (1-L_1 N_1)
(1-L_1 M_1) (1-M_2 N_2) (1-L_2 N_2) (1-L_2 M_2)]^{-1}
\cr&\times 
\left( {1\over(1-L_2 M_1
N_1) (1-L_2 M_1^2 N_2)}+{L_2 M_2 N_1^2 \over(1-L_2 M_1 N_1) (1-L_2 M_2 N_1^2)}
\right. \cr 
&\left.+{L_1^3 M_2^2 N_1 N_2 \over (1-L_1 M_2 N_1) (1-L_1^2 M_2 N_2)} 
 +{L_1 M_2 N_1 \over (1-L_1 M_2 N_1) (1-L_2 M_2 N_1^2)}\right. \cr
 &\left. +{L_1^2 M_2 N_2 \over (1-L_1 M_1 N_2) (1-L_1^2 M_2 N_2)} +
{L_1 M_1 N_2
\over (1-L_1 M_1 N_2) (1-L_2 M_1^2 N_2)}\right) \cr}\eq$$
{}From this expression, we read off the following list of  elementary couplings
(recall that the first variable is a model variable and then we write
the corresponding monomial in the grading variables):
$$\eqalign{
A_1 : M_1 N_1,\quad\quad\quad & A_2 : L_1 N_1,\quad\quad\quad ~~A_3 : L_1 M_1 \cr
B_1 : M_2 N_2,\quad\quad\quad & B_2 : L_2 N_2,\quad\quad\quad ~~B_3 : L_2 M_2 \cr
C_1 : L_2 M_1 N_1,\quad\quad & C_2 : L_1 M_2 N_1,\quad\quad ~C_3 : L_1 M_1 N_2 \cr
D_1 : L_1^2 M_2 N_2,\quad\quad & D_2 : L_2 M_1^2 N_2,\quad\quad D_3 : L_2 M_2
N_1^2 . \cr} \eq $$ 
However, not all the products of the model variables can be
linearly independent: there are linear
relations between:
$$\eqalignT{ &C_i C_j, \qquad &A_k D_k, \quad &{\rm and}\quad  A_i A_j B_k\cr
&D_i D_j,  \qquad &A_k^2 B_i B_j,  \quad &{\rm and}\quad  B_kC_k^2\cr
& C_i D_i,  \qquad & A_j B_k C_k,  \quad &{\rm and}\quad  A_k
B_j C_j\cr}\eqlabel\sprel$$ for $i,j,k$ a cyclic permutation of $1,2,3$ (and
repeated indices are not summed). (It is plain  
that the three sets of products found to be linearly related must
have the same Dynkin labels.)  A specific form of the  generating functions
amounts to a specific choice of a set of forbidden couplings among those that are
related by a linear relation.

\subsec{Technical remarks on the character method}

The character method provides a first
principle approach to the construction of tensor-product generating functions. 
This is certainly its great virtue.
However, the last two  examples indicate the
essential limitation of this method for constructing generating functions for
tensor products: the calculations are extremely complicated.  Further progress
requires the search for more powerful techniques.

It should also be clear from the last example that in general the problem
of finding a model and computing its Poincar\'e series is not trivial
even when we know the form of the generating function. Of course we
shall be interested in the inverse procedure, that is, in constructing models in
order to compute the associated generating function. This requires a systematic
procedure for computing Poincar\'e series and this will be provided
by the theory of Grobner bases which we shall discuss shortly.

%==================================================================
\newsec{Tensor-product descriptions }

%------------------------------------------------------------------

\subsec{The need for a tensor-product description}

As already mentioned, the fundamental limitation of the character method for
constructing tensor-product generating functions lies in the complication of the
intermediate calculations associated to the projection to positive powers of the
$x_i$'s variables of the different terms of the generating character products. 
The complication of these intermediate steps should be contrasted with the
relative simplicity of the resulting generating functions. This state of affair
strongly suggests that there are much more efficient ways of obtaining these
generating functions.

It is clear  that one major technical complication of the character method is that
it starts at  too fundamental a level, namely the character of the separate
representations.  As a result, we need to take care of the action of the Weyl
group: this generates many terms and the $\Omega$ projection of each term is
rather complicated. But the fact that their final sum conspires to produce a
rather simple result suggests that bypassing the use of the Weyl group would
induce  substantial simplifications.\foot{This continues to be the case when 
affine Weyl reflections are included for the fusion-rule generating functions,
with the additional complication that extra projections are required
to ensure that the representation labels are in the fundamental region
of the affine Weyl group.}

One natural way to proceed is to start from a combinatorial description of the
tensor-product rules. Such a description already
takes into account the action of the Weyl group and encodes the various
subtractions of the singular vectors.

But how do we make the connection with the generating-function approach?
The key is to find a combinatorial description which can be expressed
as a set of linear Diophantine inequalities. 

Given this set of inequalities, 
there is an algorithm, again due to 
MacMahon\foot{This is an adaptation of a
method developed by Elliot [\ref{E.B. Elliott, Quart. J. Math. {\bf 34} (1903)
388.}] for the analysis of linear Diophantine equalities and for this reason
the algorithm is often referred to as the Elliot-MacMahon method.}, for
constructing a generating function. In this  context, the generating-function
method appears as a general approach to the solutions of a system of
inequalities.  This is  particularly well illustrated in MacMahon's book
[\Mac]\foot{See in particular vol. 2 section VIII.}.  This provides then a
direct route from the  Diophantine inequalities to the generating function. 
This method is  conceptually similar to the character method, except that the
starting point is substantially closer to the end result.  
Let us illustrate MacMahon's approach
with a simple example.

%------------------------------------------------------------------

\subsec{MacMahon's theory for the solution of Diophantine inequalities}

We look for all the positive integer solutions of the inequalities  (see e.g.
[\Mac] vol. 2 no 356 p. 109)$$x_1\geq x_2\qquad x_1\geq x_3\eq$$  To impose these
constraints on a free series of the form
$$\sum X_1^{x_1}X_2^{x_2}X_3^{x_3}\eq$$
we introduce extra parameters $t$ and $s$ as follows. To take into account the first
inequality, we replace $X_1$ by $tX_1$ and $X_2$ by $t^{-1}X_2$ and project onto
positive powers of $t$. Similarly, the second inequality is taken care by the
replacements $X_1\rw sX_1$ and $X_3\rw s^{-1}X_3$ and projecting onto positive
powers of $s$. This leads to
$$\Ot\Os\; {1\over(1-tsX_1)(1-t^{-1}X_2)(1-s^{-1}X_3)}\eq$$
 whose projections read
$${1-stX_1^2X_2X_3\over(1-tsX_1)(1-sX_1X_2)(1-tX_1X_3)(1-X_1X_2X_3)}\eq$$
We then set the auxiliary variables $s,t$ equal to 1 and obtain:  
$${1-X_1^2X_2X_3\over (1-X_1)(1-X_1X_2)(1-X_1X_3)(1-X_1X_2X_3)}\eq$$
{}{}From this we read the four elementary solutions to the Diophantine inequalities
under study:
$$\eqalignD{ & \alpha_1 = (1,0,0), \qquad &\alpha_2= (1,1,0)\cr &\alpha_3= (1,0,1),\qquad
&\alpha_4= (1,1,1)
\cr}\eq$$ with the ordering
$(x_1,x_2,x_3)$ and the linear relation $$\alpha_2+\alpha_3= \alpha_1+\alpha_4\eq$$

%------------------------------------------------------------------

\subsec{{}Seeking a road from elementary couplings to generating functions}

Although the description of tensor products via linear Diophantine
equations is a more efficient route to finding the generating function
than the character one,
complications associated to the
$\Omega$ projections remain a source of technical difficulty that severely limits the
practical applicability of the method.

A more powerful approach to our problem is 
to use the techniques of computational algebra. We start with a description
of the tensor-product multiplicities as solutions to linear Diophantine
inequalities. {}From these we find directly a model for the generating
function.\foot{This is roughly the
inverse of the MacMahon's method which was originally conceived as a technique to
generate the elementary couplings and their linear relations through the construction
of the generating function.  Here, the elementary couplings and their relations are
first obtained and used as the input for the construction of the generating function.}

 In this approach, we thus work  
 with a particular model of our generating
function. But the advantage of this is that we can use Grobner-basis
techniques, described below, to find the Poincar\'e series
of the model and hence the required generating function.

%============================================================================== 

\newsec{The LR rule ($su(N)$)}

For $su(N)$ tensor products, there is a particularly
convenient description, which is that of the Littlewood-Richardson tableaux,
supplemented by the stretched-product operation (defined below) 
[\CCS].

Integrable weights in $su(N)$ can be represented by tableaux: the weight
$(\la_1,\la_2,
\cdots ,\la_{N-1})$ is associated to a left justified tableau of $N-1$ rows 
with $\la_1+\la_2+\cdots +\la_{N-1}$ boxes in the first row, $\la_1+\la_2+\cdots
+\la_{N-2}$ boxes in the second row, etc.  Equivalently, the tableau has $\la_1$
columns of 1 box, $\la_2$ columns of 2 boxes, etc. The scalar representation has
no boxes, or equivalently, any number of columns of $N$ boxes. 

The  Littlewood-Richardson
rule is a simple combinatorial description of the tensor product of two $su(N)$
representations $\la\otimes \mu$.   The second tableau ($\mu$) is filled with
numbers as follows: the first row with
$1$'s, the second row with $2$'s, etc. All the boxes with a $1$ are then added  to
the first tableau according to 
following restrictions:

\n 1) the resulting tableau must be regular: the number of 
boxes in a
given row must be smaller or equal to the number of boxes in the row 
just above;

\n 2) the resulting tableau must not contain two boxes 
marked by $1$
in the same column.

\n All the boxes marked by a $2$ are the added to
the resulting tableaux according to the above two rules (with $1$
is replaced by $2$) and the further restriction:

\n 3) in counting from right to left and top to bottom, the 
number of
$1$'s must always be greater or equal to the number of $2$'s.

\n The process is repeated with the boxes marked by a $3, 4, \cdots$, with the
additional
rule that the number of
$i$'s must always be greater or equal to the number of $i+1$'s when counted from
right to left and top to bottom.
 The resulting Littlewood-Richardson (LR) tableaux are the Young
tableaux of the irreducible representations occurring in the decomposition.

These rules can be rephrased in an algebraic way as follows. Define $n_{ij}$ to
be the number of boxes $i$ that appear in the LR tableau in the row $j$. The LR
conditions read
$$\lambda_{j-1}+\sum_{i=1}^{k-1} n_{i,j-1}-\sum_{i=1 }^{{\rm
min}(k,N-1)} n_{ij}\geq 0 \quad\quad\quad 1\leq k \leq j\leq N \quad
j\neq 1  \eqlabel\nijrang$$
and
$$\sum_{j=i}^{k} n_{i-1 \, j-1}-\sum_{j=i}^{k} n_{ij} \geq
0 \quad\quad\quad 2\leq i \leq k \leq N \quad {\rm et} \quad i\leq N-1.
\eqlabel\nijlr$$
The weight $\mu$ of the second tableau and the weight $\nu$ of the resulting LR
tableau are respectively given by 
$$\eqalign{  
  \sum_{j=i}^{N} n_{ij} & =\sum_{j=i}^{N-1} \mu_{j} \quad\quad\quad\quad
i=1,2,..., N-1 \quad ,\cr 
\nu_j -\lambda_j+\sum_{i=1}^{N-1} n_{i\, j+1} & =\sum_{i=1}^{{\rm
min}(j,N-1)} n_{ij} \quad\quad\quad\quad j=1,2,..., N-1 ~. \cr} 
\eqlabel\nijtrois$$ 
Hence, given three weights $\la,\mu$ and $\nu$, the number of positive integers
solutions $\{n_{ij}\}$ satisfying the above conditions gives the multiplicity 
$N_{\lambda
\mu}^{\quad\nu}$ of $\nu$ in the tensor product $\la\otimes \mu$.

The combined equations (\nijrang) and (\nijlr) constitute a set of linear and
homogeneous inequalities.  
As described in [\Stan], 
the Hilbert basis theorem guarantees that every solution can be
expanded in terms of the elementary solutions of these inequalities.

As explained in section 3.2 we can construct a model for the
solutions of the equations (\nijrang) and (\nijlr) by introducing
new formal variables $A_i$, $1\leq i\leq t$ where $t$ is the total
number of variables in (\nijrang) and (\nijlr). Then the subring
of $\Q[A_i; 1\leq i\leq t]$ generated by the monomials $A^\alpha$
with $\alpha$ a solution of (\nijrang) and (\nijlr) provides the
required model. This ring $R$ will be generated by a finite set of
monomials $E_j$ $1\leq j\leq s$ which we call elementary couplings
corresponding to the elementary solutions of (\nijrang)
and (\nijlr). Thus $R$ is isomorphic to $\Q[e_1,\dots,e_s]/I$ 
under the mapping $\phi: e_i\rightarrow E_i$ where $I$
is an ideal. Each element of $I$ corresponds, via the map $\phi$,
to a relation between the elementary couplings.  
As we shall see shortly, calculating a generating set
of elements of $I$ (or more particularly a Grobner basis) is the
key step in our calculations.

In the case of LR tableaux, there is a nice pictorial representation
of the model $R$. Consider the set of formal linear combinations
of 
LR tableaux with rational
coefficients. It is given a ring structure by  
defining the {\it stretched product} of two LR tableaux (denoted by $\cdot$) to
be the tableau obtained by
fusing the two tableaux and reorder the numbers in each row in increasing order. 
More algebraically, if we denote the void boxes of a LR tableau by a 0, so that
$$ n_{0j} 
=\sum_{i=j}^{N-1} \lambda_i \quad\quad\quad\quad j=1,2, ...,N-1\eq$$
we can characterize  completely a tableau by the data $\{n_{ij}\}$ with now
$i\geq 0$. 
It is clear the set of numbers $\{n_{ij}\}$ with $i\geq 0$, or equivalently,
$\{\la_i, n_{ij}\}$ with $i\geq1$, is a complete set of variables for the
description of the tensor products. 
Then, the tableau obtained by the stretched product of the
tableaux  $\{n_{ij}\}$ and  $\{ n'_{ij}\}$ is simply described by the numbers 
$\{n_{ij}+n'_{ij}\}$. 
Here is a simple example:
$$\matrix{\ST{\STrow{\bv\bv\b1}\STrow{\b1\b2\b2}\STrow{\b2\b3}\STrow{\b4}
}\cr} \cdot \matrix{\ST{\STrow{\bv\bv\b1}\STrow{\bv\b1\b2}\STrow{\bv\b2}
}\cr}  =  
\matrix{\ST{\STrow{\bv\bv\bv\bv\b1\b1}\STrow{\bv\b1\b1\b2\b2\b2}
\STrow{\bv\b2\b2\b3}\STrow{\b4} 
 }\cr}\quad  \eq $$

This ring of tableaux, then, is isomorphic to the model $R$ constructed
above and we do not distinguish between them. Thus we specify a set of
elementary couplings (i.e. a set of generators of $R$) as a set
of elementary LR Tableaux.

%------------------------------------------------------------------

\subsec{Example: the $su(2)$ case}

The complete set of inequalities for $su(2)$ variables $\{\la_1, n_{11},
n_{12}\}$ is simply 
$$\la_1 \geq n_{12} \qquad
 n_{11}\geq 0 \qquad n_{12}\geq 0\eqlabel\inedeux$$
The other weights are fixed by $$ \mu_1 =n_{11}+n_{12}\qquad 
\nu_1=\la_1+n_{11}-n_{12}\eq$$
By inspection, the elementary solutions of this set of inequalities are
$$(\la_1, n_{11}, n_{12}) = (1,0,1), \quad (1,0,0), \quad (0,1,0)\eq$$
which correspond respectively to $E_1, E_2, E_3$ in (\elesud). 
These correspond to the
following LR tableaux:
$$E_1: \quad \ST{\STrow{\bv}\STrow{\b1}}\, ,  \qquad E_2: \quad \ST{\STrow{\bv}}\,
,
\qquad E_3: \quad \ST{\STrow{\b1}}\eqlabel\tabdu$$It is also 
manifest
that there are no linear relations between these couplings.  The generating
function is thus simply:
$$G^{su(2)}= {1\over (1- E_1)(1- E_2)(1- E_3)}\eq$$
(Recall once again our convention concerning grading and model
variables: the LR tableaux in this generating function must be
replaced by the corresponding monomials in the grading variables).

%------------------------------------------------------------------

\subsec{Example: multiple tensor products in the  $su(2)$ case}

Before we turn to more complicated algebras, it is interesting to reconsider the
problem of multiple tensor products treated previously from the character method. 
Let us look for the multiplicity of the representation $\zeta$ in the triple
product
$\la\otimes \mu\otimes \nu\supset \zeta$. In the first step, the LR rule applies
as before: with $n_{11}+n_{12}= \mu_1$, we have
$\la_1\geq n_{12}$. After the first product, we re-apply the LR rule with now
$\la_1$ replaced by $\la_1+n_{11}-n_{12}$ and $n_{ij}$ replaced by $m_{ij}$ with
$m_{11}+m_{12}= \nu_1$.  The LR gives $\la_1+n_{11}-n_{12}\geq m_{12}$. The two
inequalities that defines the $su(2)$  LR basis for the quadruple product are
then
$$\la_1\geq n_{12}\qquad \la_1+n_{11}-n_{12}\geq m_{12}\qquad
n_{ij}\geq 0\qquad m_{ij}\geq 0\eqlabel\multiin$$ The elementary solutions are
then, in the order: name of the coupling, corresponding Dynkin labels and the
5-vector
$(\la_1,n_{11},n_{12},m_{11},m_{12})$,:
$$\eqalignT{
 &E_1: &(1)\otimes(1)\otimes(0)\supset (0)\qquad& (1,0,1,0,0)\cr  
 &E_2: &(1)\otimes(0)\otimes(1)\supset (0)\qquad &(1,0,0,0,1)\cr 
 &E_3: &(1)\otimes(0)\otimes(0)\supset (1)\qquad &(1,0,0,0,0)\cr 
 &E_4: &(0)\otimes(1)\otimes(1)\supset (0)\qquad &(0,1,0,0,1)\cr 
 &E_5: &(0)\otimes(1)\otimes(0)\supset (1)\qquad &(0,1,0,0,0)\cr 
 &E_6: &(0)\otimes(0)\otimes(1)\supset (1)\qquad &(0,0,0,1,0)\cr}\eqlabel\mutipo$$
The linear relation, whose existence was signalled by the character method, is
$$E_3E_4= E_2E_5: (1,1,0,0,1), \qquad \not= E_1E_6: (1,0,1,1,0)\eq$$
 Choosing to forbid the product $E_3E_4$, the
generating function can be written in the form
$$\eqalign{ G &= {1-E_3E_4\over
(1-E_1)(1-E_2)(1-E_3)(1-E_4)(1-E_5)(1-E_6)}\cr
&= \left(\prod_{i=1,2,5,6}{1\over 1-E_i}\right)\left({1\over 1-E_3}+{E_4\over
1-E_4}\right)\cr}\eqlabel\multi$$
The latter form makes manifest the absence of $E_3E_4$.

We could represent the  elementary couplings in terms of tableaux, where the boxes
with 1 refers to the $\mu$ tableau and those with 2 originates from the $\nu$
tableau.  (Warning: the resulting tableaux describing the four-products are not
 necessarily LR tableaux.) Hence, $n_{1j}$ gives the number of 1 in row $j$ of the
composed tableau while $m_{1k}$ gives the number of 2 in row $k$.  The elementary
tableaux are 
$$\eqalignT{ &E_1:\quad \ST{\STrow{\bv}\STrow{\b1}}  \qquad &E_2: \quad
\ST{\STrow{\bv}\STrow{\b2}}\, ,
& E_3: \quad \ST{\STrow{\bv}}\cr
&E_4:\quad \ST{\STrow{\b1}\STrow{\b2}}  \qquad &E_5: \quad
\ST{\STrow{\b1}}\, ,
& E_6: \quad \ST{\STrow{\b2}}\cr}\eq$$
{}From this representation, the relation reads
$$E_3E_4= E_2E_5: \ST{\STrow{\bv\b1}\STrow{\b2}}, \qquad \not= E_1E_6:
\ST{\STrow{\bv\b2}\STrow{\b1}}\eq$$

 It is worth pointing out here the relation
between the multiplicities of the
$su(2)$ triple tensor products and $su(3)$ weight multiplicities in
$su(3)$ irreducible representations. To make this relation explicit, we first
consider void boxes to be filled with 0 and then reshuffle all filling numbers by
1, i.e.,
$$\ST{\STrow{\bv\b2}\STrow{\b1}} \rw \ST{\STrow{\b1\b3}\STrow{\b2}}\eq$$
The multiplicity of the product $\la\otimes \mu\otimes \nu\supset \zeta$ is then
seen to be the number of tableaux filled by $\la_1$ 1's , $\mu_1$ 2's, $\nu_1$
3's (the tableau has thus a total number of $\la_1+\mu_1+\nu_1$ boxes)
with the restriction that there should be 
$\zeta_1$ columns of 1 box. Since by construction, the numbers are strictly
increasing in each column from top to bottom  and non-decreasing in each row from
left to right, these tableaux are nothing but $su(3)$ semistandard tableaux. The
resulting tableau has
$su(3)$  weight\foot{In a $su(N)$ semistandard tableau, a box filled by $i$ has
weight
$-\om_{i-1}+\om_i$, with
$\om_0=\om_N=0)$.}
$$\la_1\om_1 + \mu_1(-\om_1+\om_2)-\nu_1\om_2\eq$$ and row lengths:
$(\la_1+\mu_1+\nu_1+\zeta_1)/2$ and $(\la_1+\mu_1+\nu_1-\zeta_1)/2$.  In terms of
Dynkin labels, the number of such semistandard tableaux is exactly the multiplicity
of the
$su(3)$ weight
$(\la_1-\mu_1, \mu_1-\nu_1) $ in the representation of highest weight $(\zeta_1,
(\la_1+\mu_1+\nu_1-\zeta_1)/2)$. For instance, the $su(2)$ product
$(1)\otimes(1)\otimes(1)\supset (1)$ has multiplicity 2, corresponding to the
products
$E_3E_4$ (forbidding $E_2E_5$) and $E_1E_6$, or equivalently, to the two tableaux
$$\ST{\STrow{\bv\b1}\STrow{\b2}}, \qquad 
\ST{\STrow{\bv\b2}\STrow{\b1}}\eq$$ and this is exactly the multiplicity of the
$su(3)$ weight $(0,0)$ in the highest-weight representation $(1,1)$, whose two
semistandard tableaux are $$\ST{\STrow{\b1\b2}\STrow{\b3}}, \qquad 
\ST{\STrow{\b1\b3}\STrow{\b2}}\eq$$
Equivalently, since the multiplicity of the $su(2)$ product $\la\otimes \mu\otimes
\nu\supset \zeta$ is the same as
$\la\otimes \mu\otimes
\nu\otimes \zeta \supset (0)$, by filling the boxes of $\zeta$ with 4's, one can
relate the resulting multiplicity to a $su(4)$ weight multiplicity in an
irreducible representation of two equal rows.  More precisely, the resulting
tableaux have  $\la_1$ 1's,  $\mu_1$ 2's, $\nu_1$
3's and $\zeta_1$ 4's, which corresponds to a weight 
$$\la_1\om_1 + \mu_1(-\om_1+\om_2)+\nu_1(-\om_2+\om_3)-\zeta_1\om_3\eq$$
and the corresponding irreducible representation has weight $(0,n,0,0)$ with 
$$ 2n = \la_1+\mu_1+\nu_1+\zeta_1\eq$$
More generally, the multiplicity of the multiple $N\geq 3$ $su(2)$ product 
$$\bigotimes_{i=1}^N \la^{(i)} \supset (0)\eq$$ is equal to the multiplicity of
the
$su(N)$ weight $\Lambda'$ in the highest-weight representation
$\Lambda$, with
$$\Lambda'= \sum_{i=1}^N
\la_i^{(i)}(-\om_{i-1}+\om_i)\qquad \Lambda=(0,\frac12\sum_{i=1}^N
\la_i^{(i)},0,\cdots,0)\eq$$  (with
$\om_0=\om_N=0)$.

%------------------------------------------------------------------

\subsec{Example: the $su(3)$ case}

Let us return to standard triple tensor products and turn to $su(3)$.
The LR conditions for $su(3)$ are
$$
\eqalign{ \la_1  &\geq n_{12}\cr
 \la_2 &\geq n_{13}\cr  \la_2+n_{12} &\geq
n_{13}+n_{23}\cr
 n_{11}&\geq n_{22} \cr n_{11}+n_{12}&\geq n_{22}+n_{23}\cr}
\eqlabel\inee$$
The other weights are given by
$$\eqalign{ \mu_1 &= n_{11}+n_{12}+n_{13}-n_{22}-n_{23}\cr
\mu_2 &= n_{22}+n_{23}\cr
\nu_1 &= \la_1+n_{11}-n_{12}-n_{22}\cr
\nu_2 &= \la_2+n_{12}+n_{22}-n_{13}-n_{23}\cr}\eqlabel\other$$

The elementary solutions of the set of inequalities (\inee) are again easily
found by inspection
\foot{In section 7, we present a
 way to generate the elementary solutions 
 of a system of Diophantine inequalities
 starting from the construction of a vector basis. This method
 appears to be avoid some of the computational complexity of
 the MacMahon method and in many cases 
 the calculations can be done by hand.}   
 and they are given
by the following set of numbers:
$ (\la_1,
\la_2, n_{11}, n_{12}, n_{13}, n_{22}, n_{23})$:
$$\eqalign{
E_1: \quad (1,0)\otimes(0,1)\supset(0,0): & \quad (1,0,0,1,0,0,1) \cr 
E_2: \quad (1,0)\otimes(0,0)\supset(1,0): & \quad (1,0,0,0,0,0,0) \cr 
E_3: \quad (0,0)\otimes(1,0)\supset(1,0): & \quad (0,0,1,0,0,0,0) \cr 
E_4: \quad (0,1)\otimes(1,0)\supset(0,0): & \quad (0,1,0,0,1,0,0) \cr 
E_5: \quad (0,1)\otimes(0,0)\supset(0,1): & \quad (0,1,0,0,0,0,0) \cr 
E_6: \quad (0,0)\otimes(0,1)\supset(0,1): & \quad (0,0,1,0,0,1,0) \cr 
E_7: \quad (1,0)\otimes(1,0)\supset(0,1): & \quad (1,0,0,1,0,0,0) \cr 
E_8: \quad (0,1)\otimes(0,1)\supset(1,0): & \quad (0,1,1,0,0,0,1) \cr }\eq$$ The
corresponding tableaux are
 $$\eqalign{ 
  E_1 &: \quad\ST{\STrow{\bv}\STrow{\b1}\STrow{\b2} }, \qquad  
 E_2: \quad\ST{\STrow{\bv}}, \qquad
 E_3: \quad\ST{\STrow{\b1}}, \qquad
 E_4: \quad\ST{\STrow{\bv}\STrow{\bv}\STrow{\b1}} \cr
   E_5 & : \quad\ST{\STrow{\bv}\STrow{\bv}}, \qquad
  E_6 : \quad\ST{\STrow{\b1}\STrow{\b2}}, \qquad
  E_7 : \quad\ST{\STrow{\bv}\STrow{\b1}}, \qquad
  E_8 :\quad \ST{\STrow{\bv\b1}\STrow{\bv}\STrow{\b2}}.
 \cr} \eq$$
If we use the 7-component vector as a description of an elementary coupling,
we see that there is one relation 
$$E_1E_3E_5= E_7E_8  : (1,1,1,1,0,0,1)\eq$$
This is confirmed by the construction of the corresponding LR tableaux:
$$E_1E_3E_5: \ST{\STrow{\bv}\STrow{\b1}\STrow{\b2} }\cdot
\ST{\STrow{\b1}}\cdot \ST{\STrow{\bv}\STrow{\bv}}=
\ST{\STrow{\bv\bv\b1}\STrow{\bv\b1}\STrow{\b2}} \eq$$ and 
$$ E_7E_8 : 
\ST{\STrow{\bv}\STrow{\b1}}\cdot 
\ST{\STrow{\bv\b1}\STrow{\bv}\STrow{\b2}}
= \ST{\STrow{\bv\bv\b1}\STrow{\bv\b1}\STrow{\b2}}\eq$$
The generating function can thus be constructed by either forbidding $E_1E_3E_5$
or $E_7E_8$.  That yields directly the first or the third expression of
(\fgsutr) respectively.\foot{In this basis, $E_2E_4E_6$ is an independent
product.}

%------------------------------------------------------------------

\subsec{Example: the $su(4)$ case}

To the $su(4)$ LR known conditions are
$$\eqalign{
\la_1& \geq n_{12}\cr
\la_2& \geq n_{13}\cr
\la_2+n_{12}& \geq n_{13}+n_{23}\cr
\la_3& \geq n_{14}\cr
\la_3+n_{13}& \geq n_{14}+n_{24}\cr
\la_3+n_{13}+n_{23}& \geq n_{14}+n_{24}+  n_{34}\cr
n_{11}& \geq n_{22}\cr
n_{11}+n_{12}& \geq n_{22}+ n_{23}\cr
n_{11}+n_{12}+n_{13}& \geq n_{22}+ n_{23}+n_{24}\cr
n_{22}& \geq n_{33}\cr
n_{22}+n_{23}& \geq n_{33}+ n_{34}\cr}\eqlabel\inequatre$$

The tensor-product elementary couplings, that is, the elementary solutions to
these inequalities are best written directly in terms of LR tableaux:
$$\eqalign{
A_1 &:\ST{\STrow{\b1}\STrow{\b2}\STrow{\b3} }, \quad  
 A_2:\ST{\STrow{\bv}\STrow{\bv}\STrow{\bv}\STrow{\b1}}, \quad
 A_3:\ST{\STrow{\bv}},\quad
B_1 :\ST{\STrow{\b1}\STrow{\b2} }, \quad  
 B_2:\ST{\STrow{\bv}\STrow{\bv}\STrow{\b1}\STrow{\b2}}, \quad
 B_3:\ST{\STrow{\bv}\STrow{\bv}},\cr
C_1 &:\ST{\STrow{\b1} }, \quad  
 C_2:\ST{\STrow{\bv}\STrow{\b1}\STrow{\b2}\STrow{\b3}}, \quad
 C_3:\ST{\STrow{\bv}\STrow{\bv}\STrow{\bv}},\quad
D'_1 :\ST{\STrow{\bv}\STrow{\bv}\STrow{\b1} }, \quad  
 D'_2:\ST{\STrow{\bv}\STrow{\b1}}, \quad
 D'_3:\ST{\STrow{\bv}\STrow{\b1}\STrow{\b2}},\cr}\eqlabel\elequ$$
together with
$$\eqalign{
D_1 &:\ST{\STrow{\bv\b1}\STrow{\bv}\STrow{\b2}\STrow{\b3} }, \quad 
D_2 :\ST{\STrow{\bv\b1}\STrow{\bv\b2}\STrow{\bv}\STrow{\b3} }, \quad 
D_3 :\ST{\STrow{\bv\b1}\STrow{\bv}\STrow{\bv}\STrow{\b2} }, \cr 
E_1 &:\ST{\STrow{\bv\bv}\STrow{\bv\b1}\STrow{\bv}\STrow{\b2} }, \quad 
E_2 :\ST{\STrow{\bv\b1}\STrow{\bv}\STrow{\b2} }, \quad 
E_3 :\ST{\STrow{\bv\b1}\STrow{\bv\b2}\STrow{\b1}\STrow{\b3} }, \quad
\cr}\eqlabel\elequu$$ The Dynkin-label transcription of the elementary couplings
reads
$$\eqalignD{
&A_1:(0,0,0)\otimes(0,0,1)\supset(0,0,1)\qquad
&D_1':(0,1,0)\otimes(1,0,0)\supset(0,0,1)\cr
&A_2:(0,0,1)\otimes(1,0,0)\supset(0,0,0)\qquad
&D_2':(1,0,0)\otimes(1,0,0)\supset(0,1,0)\cr
&A_3:(1,0,0)\otimes(0,0,0)\supset(1,0,0) \qquad
&D_3':(1,0,0)\otimes(0,1,0)\supset(0,0,1) \cr 
&B_1:(0,0,0)\otimes(0,1,0)\supset(0,1,0)\qquad
&D_1:(0,1,0)\otimes(0,0,1)\supset(1,0,0)\cr 
&B_2:(0,1,0)\otimes(0,1,0)\supset(0,0,0)\qquad 
&D_2:(0,0,1)\otimes(0,0,1)\supset(0,1,0)\cr 
&B_3:(0,1,0)\otimes(0,0,0)\supset(0,1,0)\qquad
&D_3:(0,0,1)\otimes(0,1,0)\supset(1,0,0) \cr 
&C_1:(0,0,0)\otimes(1,0,0)\supset(1,0,0)\qquad 
&E_1:(1,0,1)\otimes(0,1,0)\supset(0,1,0)\cr 
&C_2:(1,0,0)\otimes(0,0,1)\supset(0,0,0)	\qquad
&E_2:(0,1,0)\otimes(0,1,0)\supset(1,0,1)\cr  
&C_3:(0,0,1)\otimes(0,0,0)\supset(0,0,1)\qquad
&E_3:(0,1,0)\otimes(1,0,1)\supset(0,1,0)\cr}\eq$$   
For $su(4)$, there is a large number of linear relations: in fact there are 15
relations [\ref{R.T Sharp and D. Lee, Revista Mexicana de Fisica {\bf 20} (1971)
203.}\refname\SL,\CCS] :
$$ \eqalignT{ &D_j^{'}  D_ k = C_i E_i \qquad &D_j D_k^{'}  = B_i C_j C_k  \qquad
&E_i E_j  = B_k D_k D_k^{'} \cr
&D_i E_i  = C_j B_k D_k \qquad
&D_i^{'}E_i   = B_j D_j^{'} C_k \qquad &\quad {} \cr}
\eqlabel\mmmhb$$
with $ i,j,k$ a cyclic permutation of $1,2,3$.

 To construct the generating function, we need to select forbidden couplings. 
It turns out that when there are more that one relation, complications
may arise.  We must ensure that the selected forbidden
couplings are complete, which means that no further (usually higher-order)
relations are required for a unique decomposition of a given coupling.  How
do we select a set of complete compatible forbidden couplings? A technique that is
tailor-made for dealing with problems of that type is that of Grobner bases. 
This will be introduced in the next section. At this point, we simply indicate a
complete choice of forbidden couplings, namely   
$\{E_iE_j, D'_iE_i, D_iE_i, D_jD'_i,D'_jD_i\}$.  This yields then a model for
the generating function, which then reads [\SL,\CCS] :
$$\eqalign{
G^{{\rm su}(4)}=& (
\prod_{i=1}^3~ \tilde{A}_i \tilde{B}_i \tilde{C}_i ) (
\tilde{D}_1' \tilde{D}_2' \tilde{D}_3'
+E_1 \tilde{E_1} \tilde{D}_2' \tilde{D}_3'
+D_3 \tilde{D}_3 \tilde{D}_3'\tilde{E}_1 \cr 
&+D_2 \tilde{D}_2 \tilde{D}_3 \tilde{E}_1
+D_1 \tilde{D}_1 \tilde{D}_2 \tilde{D}_3 
+E_3 \tilde{E}_3 \tilde{D}_1 \tilde{D}_2
+D_1' \tilde{D}_1' \tilde{D}_1 \tilde{E}_3  \cr
&+D_2' E_3 \tilde{D}_2' \tilde{E}_3 \tilde{D}_1'
+E_2 \tilde{E}_2\tilde{D}_1' \tilde{D}_3' 
+E_2 D_1 \tilde{E}_2 \tilde{D}_1 \tilde{D}_1' 
+ E_2 D_3 \tilde{E}_2 \tilde{D}_3 \tilde{D}_3' \cr
&+D_1 D_3 E_2  \tilde{D}_1 \tilde{D}_3 \tilde{E}_2 
+D_2 D_2' \tilde{D}_2 \tilde{D}_2' \tilde{E}_1 
+D_2 D_2' E_3  \tilde{D}_2 \tilde{D}_2' \tilde{E}_3 ). \cr}
\eqlabel\fcsqr$$
where
$$\tilde{M}_i =(1-M_i)^{-1} .\eq$$
%----------------------------------------------------------------------------

\subsec{A remark on the reduction of grading variables and higher multiplicities}

In this section, we would like to stress the fact that a
generating function built from a complete basis (or a complete description of
tensor products) has `multiplicity coefficients' all either 0 or 1: the
`coupling' either exists or not. Higher multiplicities can be generated only  
 after the number of grading
variables has been reduced (that is, some grading variables have been set equal
to 1).  To be explicit, suppose that we consider the generating function for the
$su(N)$ tensor products starting from a LR description.  This is a system of
inequalities for the variables
$\{\la_i,n_{jk}\}$ with $i,j=1,\cdots, N-1$ and $j\leq k=\leq N$.
Introducing the grading variables $\{L_i, N_{jk}\}$, the generating function
reads:
$$G= \sum C_{\{\la_i,n_{jk}\}} L_i^{\la_i}N_{jk}^{n_{jk}}\eq$$
and the coefficient $C_{\{\la_i,n_{jk}\}}$ is either 1, if the solution exits, or
0, if the solution does not exist.  In other words, when expressed in terms of
a complete set of variables, the multiplicity
 is 0 or 1. Larger values for the multiplicity can only result from the reduction
in the number of grading variables.  Typically, we consider the reduction from the
set
$\{\la_i,n_{jk}\}$ to the set
$\{\la_i,\mu_i, \nu_i\}$. The multiplicity, which now has the interpretation of
a tensor-product coefficient, is then no longer trivially 0 or 1 (except for
$su(2)$ where this changes of variables does not induce a reduction).

%============================================================================== 
\newsec{Diophantine inequalities: elementary couplings, relations and Grobner
bases}

We will introduce the idea of the Grobner basis via a simple
example.\foot{For an elementary introduction to Grobner bases, see for instance
 [\ref{R. Froberg, {\it An introduction to Grobner bases}, Wiley, New York
 1997.}].} 
Suppose we know that
$R=Q[x,y,z,t]/I$ where
$I=(xy-t,zy-t)$, with an $\N^2$ grading given by $(1,0), \,(0,1), \,(1,0)$ and 
$(1,1)$ for $x,\, y,\,z$ and $t$, is a model for a generating function. Writing
$\bar x = x+I$ and similarly for the other variables, we have in $R$
that $\bar x \bar y = \bar t$ and $\bar z\bar y =\bar t$. These two expressions
give two {\it re-write rules} : 
%* correction $xy \mapsto z$ -> $xy \mapsto t$:
$xy \mapsto t$
and $zy \mapsto t$. These rules can be used to simplify any 
monomial. The aim is to find a re-write rule which, when iterated, produces
unique representatives for the classes of $I$. If this is the case, then a
vector space basis of $R$ would consist of terms of the form $m+I$ with
$m$ a monomial which is not divisible by any of the left-hand sides of the
rewrite rules. 
 
In the example above, if we had  `good' rewrite rules then a basis for
$R$ would be represented by monomials not containing $xy$ or $zy$, i.e.
monomials of the form either $y^at^b$ or
$x^az^bt^c$. The generating function which counts there monomials, taking
into account the grading and potential over-counting, is:
$$
{1\over {(1-AB)}}\left( {B\over{1-B}}  +
{{1}\over{(1-A)^2}}\right),
\eq$$
The exponent of $A$ carries the first grading index and $B$ 
the second.\foot{In more
details: with the specified grading, $y^at^b$ corresponds 
to the term $B^a (AB)^b$
which is generated by
$[(1-AB)(1-B)]^{-1}$ while $x^az^bt^c$ corresponds to $A^aA^b(AB)^c$ which is
generated by  $[(1-AB)(1-A)^2]^{-1}$.  Since the constant term would be counted twice
if we simply add these two pieces, we multiply the first one by $B$.}

However this generating function is not correct. It contains the
term $2A^2B$ corresponding to the 2 monomials  $xt$ and
$zt$. But the polynomial $ z(xy-t) - x(zy-t) = xt-zt$ is also in $I$
and hence in $R$ we have $\bar x\bar t = \bar z\bar t$ and so
the space of grade $(2,1)$ has dimension 1 rather than 2. This
problem can also be seen as a problem with the re-write rules.
If we start with $xyz$ then we can use the first re-write
rule: $xyz \mapsto tz$ or the second: $xyz \mapsto xt$. We 
cannot apply any further re-write rules and so this set of
re-write rules does not produce a unique representative. The
solution is to include the rule $xt \mapsto zt$. This gives
a set of 3 rules: $xy \mapsto t$, $ zy \mapsto t$ and $
xt \mapsto zt$. It turns out that this is a `good' set
 and so a basis for $R$ is given by (the
classes of) monomials of the form $y^at^b$, $x^az^b$ and
$z^at^b$ which gives the generating function:
$$
{1\over {(1-AB)(1-B)}} + {A\over{(1-A)^2}} + {A\over{(1-A)(1-AB)}}
\eq$$

The set of `good' generators, $xy-t,zy-t,xt-zt$ we have found
for $I$ is known as a {\it Grobner basis} [\ref{B. Buchberger, 
Applications of Grobner basis in non-linear computational geometry
in {\it Trends in Computer Algebra, Lecture Notes in Computer Science}
{\bf 296} ed R Jansen (Berlin: Springer) (1989) 52-80.}\refname\Buch].

The general procedure for constructing a Grobner basis given a 
set of generating polynomials is as follows. First choose
a {\it term ordering}, which is an ordering on monomials with
the property that any chain $m_1 > m_2 > \dots $ has finite
length. For example we can order the variables
by $x>y>z>t$ and then order all
monomials by the corresponding lexicographic (dictionary) order, for
example:
$x^2y > xyz > y^3 $.
 For each
generator of our ideal
$I$, select the monomial which is highest with respect to the given
term ordering. This is then the term which appears on the
left of the re-write rule. The lexicographic ordering gives the
first two re-write rules of our example: $xy \mapsto t$ and
$zy \mapsto t$. Next, for each pair of leading terms find the
lowest common multiple and simplify it in the two possible ways.
In this case there is only one pair of leading terms and
the lowest common multiple is $xyz$ which simplifies to
%z made wording clearer
$xt$ and $yt$. Continue to apply the re-write rules 
until the terms  do not simplify further. If the resulting pair of terms
are the same, then proceed to the next pair of leading
terms, otherwise add a new re-write rule. In this case we add 
$xt\mapsto yt$. Proceed until no pair of leading terms gives
a new rule. This is the case for the rules we now have. For example
the two rules $xy\mapsto t$ and $xt \mapsto zt$ appears to give
a new rule by simplifying $ xyt$ to both $t^2$ and $yzt$. 
%+ small rewording:
However the
second term can be further reduced to $ t^2$ and so no new rule
is required. 
Improvements on this basic algorithm, known as the
Buchberger [\Buch] algorithm, mean that it is now feasible to
find Grobner bases for quite large sets of 
generating polynomials.\foot{The web pages of the
computer-algebra information network at http://cand.can.nl/CAIN
contain information about many of the programs currently
available.}

Although it is not clear from this example, the technique of Grobner bases is a
very versatile tool for performing explicit calculations. 
We end this section with an illustrative example  relevant to our discussion
of tensor-product generating functions.

Consider a set of linear Diophantine equations:
$$
M\alpha = 0,\qquad \alpha \geq 0
\eq$$
with $M$ an integer matrix and $\alpha$ a vector of non-negative integers.
We would like to construct a generating function for the solutions to
this set of equations:
$$
\sum_{\alpha} x^\alpha.\eq$$
A non-trivial example is given by
 the Diophantine equations that describe
 a $3\times 3$ magic square:
$$\pmatrix { a&b&c\cr d&e&f\cr g&h&i}\eq
$$
with non-negative entries and equal row and column sums. The magic square
condition
%à add
(the sum of each row and each column is the same, say equal to $t$) gives the
following set of equations:
$$
\eqalign{
a+b+c &= t\cr
d+e+f &= t\cr
g+h+i &= t\cr
a+d+g &= t\cr
b+e+h &= t\cr
c+f+i &= t\cr
}
\eq$$
With $\alpha$ standing for the column vector with entries $(a,b,c,d,e,f,g,h,i,t)$,
the matrix
$M$ reads
$$M= \pmatrix{1&1&1&0&0&0&0&0&0&-1\cr
0&0&0&1&1&1&0&0&0&-1\cr
0&0&0&0&0&0&1&1&1&-1\cr
1&0&0&1&0&0&1&0&0&-1\cr
0&1&0&0&1&0&0&1&0&-1\cr
0&0&1&0&0&1&0&0&1&-1\cr
}\eq$$

It can be show than any solution of $M\alpha=0$ is a linear combination
(with non-negative coefficients!) of a finite number of basic solutions
$\alpha_1,\cdots,\alpha_s$ (see for example [\Stan]). 
Moreover there is a straightforward
algorithm for finding this basic set of solutions [\ref{G. Huet,
{\it An algorithm to generate the basis of solutions to 
homogeneous Diophantine equations}, Information Processing Lett.
{\bf 7} (1978) 144-7.}]. In this case
we have the following set:
$$\eqalign{
  \alpha_1 &= (0  , 0 ,  1 ,  0  , 1 ,  0 ,  1 ,  0 ,  0  , 1 )\cr
   \alpha_2 &= (0  , 1 ,  0 ,  0 ,  0 ,  1 ,  1 ,  0 ,  0  , 1 )\cr
   \alpha_3 &= (0  , 0 ,  1 ,  1 ,  0 ,  0 ,  0 ,  1 ,  0 ,  1 )\cr
   \alpha_4 &= (1  , 0 ,  0  , 0 ,  0 ,  1 ,  0 ,  1 ,  0 ,  1 )\cr
   \alpha_5 &= (0  , 1 ,  0 ,  1 ,  0 ,  0 ,  0 ,  0 ,  1 ,  1 )\cr
   \alpha_6 &= (1  , 0 ,  0 ,  0 , 1 ,  0 ,  0 ,  0  , 1  , 1)\cr}
\eq$$

We shall use $A,B,\dots,T$ to denote the ``grading variables''
of this example so that the exponent of $A$ carries
the value of $a$ and so on.
A model for the generating function is given by the
subring $S$ of $\Q[A,B,C,D,E,F,G, H,I,T]$ generated by monomials
corresponding to the 6 elementary solutions, 
$$
\eqalignT{
E_1&=CEGT,\qquad 
 E_2&=BFGT,\qquad 
 E_3&=CDHT,\cr
 E_4&=AFHT,\qquad 
 E_5&=BDIT,\qquad
 E_6&=AEIT\cr}
\eq$$
The monomials in $S$ correspond to magic squares. For example
$E_1^2E_4E_6 = A^2C^2E^3FG^2HIT^4\in S$ corresponds to a square
with row and column sums equal to 4:
$$
\pmatrix{ 2&0&2\cr 0&3&1\cr 2&1&1\cr}.
\eq$$
Note that in this example it is convenient
to construct our model as a subring
 of the ring of grading variables. 
Thus each ``elementary coupling'' $E_i$ is actually
equal to the corresponding monomial in the grading
variables.

However, there are relations between these generators
and so it is not immediately clear how to construct the Poincar\'e series
for $S$. What we require is an isomorphism of $S$ with $R = \Q[e_1,\dots,e_6]/I$
such that $e_i\mapsto E_i$, $i=1,\dots,6$ and such that we have a Grobner
basis of the ideal $I$ (the `ideal of relations').

Fortunately, such an isomorphism is easily constructed using Grobner-basis
methods. Introduce the ring $\Q[A,B,C,D,E,F,G,H,I,T,E_1,\dots,E_6]$
with the lexicographic ordering $$A>B>C>D>E>F>G>H>I>T>e_1>\dots > e_6\eq$$
Let $J$ be the ideal generated by $E_1-e_1,\dots, E_6-e_6$. This is
not necessarily a Grobner basis with respect to this term ordering. Let $G$ be the Grobner basis
for $J$ with the given ordering. Then it can be shown [\Buch] that $G\cap \Q[e_1,\dots,e_6]$
is a Grobner basis for the ideal of relations $I$ which we require. 
In this case $G$ is quite large, but its intersection with $\Q[e_1,\dots,e_6]$
is $e_1e_4e_5-e_2e_3e_6$. The corresponding relation in $R$ is
 $E_1E_4E_5-E_2E_3E_6$
and it is straightforward to verify that these two
terms do indeed give the same magic square, so that indeed
we have found a relation between the generators of $R$. The Poincar\'e series
for $\Q[e_1,\dots,e_6]/I$ is easily computed:
$$ \eqalign{
{1\over(1-E_2)(1-E_3)(1-E_6)} &\left(
{1\over{(1-E_1)(1-E_4)}} \right.\cr  & \, +
\left.{{E_5}\over{(1-E_1)(1-E_5)}} +{{E_4E_5}\over{(1-E_4)(1-E_5)}}\right)\cr}
\eqlabel\majgen
$$

If we do not need to keep track of the row and column sums we can set
$T=1$ in this generating function. Similarly, if we simply want a
generating function for the number of magic squares with given row
and column sums, we can set all variables except $T$ equal to 1,
which yields:
$$
{{1+T+T^2}\over{(1-T)^5}}.
\eqlabel\teqn$$
It is perhaps worth pointing out that the model we constructed
above is still a model for ($\teqn$) if we change  
the grading by keeping just the last component of the grading
vector. Now, however, the grading variable is simply $T$ and
so we cannot identify the grading and model variables.
Thus to obtain ($\teqn$) from ($\majgen$) we once again
must use our convention that model variables are replaced
by the corresponding grading variables. In this case
this  corresponds precisely to setting all the variables
except $T$ equal to 1.

We have made use of the techniques described in this section to 
verify the results we have quoted previously.

%============================================================================== 

\newsec{The Berenstein-Zelevinsky basis}

%------------------------------------------------------------------

\subsec{Generalities}

The presentation of the $su(2,3,4)$ cases has made clear the usefulness of a
basis, or more precisely, a re-expression of the tensor-product calculation in
terms of Diophantine  equations.  The Littlewood-Richardson algorithm yields a
set of such equations only for $su(N)$.  Fortunately, 
Berenstein and Zelevinsky [\ref{ A.D. Berenstein and
A.V.  Zelevinsky, J. Geom. Phys. {\bf 5} (1989) 453.}\refname\BZ] have expressed
the solution of the multiplicity of a given tensor product as a counting problem
for the number of integral points in a convex polytope.  For a given algebra, the
polytope is formulated in terms of a characteristic set of inequalities.  These
inequalities can be interpreted as defining a tensor-product basis. For
$su(N)$, this basis reduces  to the LR set of inequalities.  For the other
classical algebras but $sp(4)$, the proposed basis is a conjecture. Hence, in
order to analyse our next example,
$sp(4)$, we first need to present the corresponding Berenstein-Zelevinsky
inequalities.

%------------------------------------------------------------------

\subsec{The BZ $sp(4)$ basis}

The combinatorial description of tensor products for $sp(4)$ is not as simple as
in the $su(N)$ case: 
a standard LR product must be supplemented by a division
operation and modification rules [\ref{G.R.E. Black, R.C  King and B.G.  Wylbourne, J. Phys. A: Math.
Gen. {\bf 16} (1983) 1555.}\refname\BGW]. Various natural trials for the search
of a linear basis  failed (and for instance, a stretched product making
manifest the relations between the elementary couplings cannot be defined). 

Given the BZ set of inequalities, the natural way to proceed, as just mentioned,
 is to interpret these as the appropriate basis for the description of the
tensor products. These inequalities are as follows:\foot{Our notation is different
from that used in [\BZ].  The relation is $r_1=m_1,\, r_2=m_2,\,
p_{12}=m_{12},\,q_{12}=m_{12}^\dagger$.}
$$
\eqalign{\lambda_1 & \geq p_{12} \cr
\lambda_2 & \geq r_{1}/2 \cr \lambda_2 & \geq r_{1}/2+q_{12}-p_{12}
\cr \lambda_2 & \geq r_{2}/2+q_{12}-p_{12} \cr 
\mu_1 & \geq q_{12} \cr
\mu_1 & \geq q_{12}+r_1-r_{2} \cr
\mu_1 & \geq p_{12}+r_1-r_{2} \cr
\mu_2 & \geq r_{2}/2 \cr
\nu_1 & = r_2-r_1-2p_{12}+\la_1+\mu_1\cr
\nu_2 & = p_{12}-q_{12}-r_2+\la_2+\mu_2\cr}
 \eqlabel\bzspq $$
The $sp(4)$ tensor product coefficient  ${\cal
N}_{\lambda\mu\nu}$ is thus given by the number of solutions of the above system
with
 $r_1,r_2 \in 2~{\Z}_{+}$ et $p_{12},q_{12} \in
{\Z}_{+}$.

A proper set of variables for a complete description of a particular
tensor-product coupling is thus
$\{\la_1,\la_2,\mu_1,\mu_2, r_1,r_2,p_{12},q_{12}\}$. We give the list of
elementary couplings, adding to each coupling the corresponding four-vector $[
r_1,r_2,p_{12},q_{12}]$:
$$\eqalignT{
&A_1:  &(0,0)\otimes(1,0)\supset(1,0)\quad [0,0,0,0]\cr
&A_2:  &(1,0)\otimes(0,0)\supset(1,0)\quad [0,0,0,0]\cr
&A_3:  &(1,0)\otimes(1,0)\supset(0,0)\quad [0,0,1,1]\cr
&B_1:  &(0,0)\otimes(0,1)\supset(0,1)\quad [0,0,0,0]\cr
&B_2:  &(0,1)\otimes(0,0)\supset(0,1)\quad [0,0,0,0]\cr
&B_3:  &(0,1)\otimes(0,1)\supset(0,0)\quad [2,2,0,0]\cr
&C_1:  &(0,1)\otimes(1,0)\supset(1,0)\quad [0,0,0,1]\cr
&C_2:  &(1,0)\otimes(0,1)\supset(1,0)\quad [0,2,1,0]\cr
&C_3:  &(1,0)\otimes(1,0)\supset(0,1)\quad [0,0,1,0]\cr
&D_1:  &(2,0)\otimes(0,1)\supset(0,1)\quad [0,2,2,0]\cr
&D_2:  &(0,1)\otimes(2,0)\supset(0,1)\quad [2,0,0,0]\cr
&D_3:  &(0,1)\otimes(0,1)\supset(2,0)\quad [0,2,0,0]\cr}\eqlabel\spele$$

The unspecified linear relations mentioned in (\sprel) can now be
obtained.   To find those products that are really equal in the
present basis, we only need to compare their corresponding sets of four-vectors
$[r_1,r_2,p_{12},q_{12}]$ (which are additive in products of
couplings).  We thus find for instance that 
$$C_1C_2= A_3D_3:\; [0,2,1,1]  \qquad \not= A_1A_2B_3 :\; [2,2,0,0]\eq$$
Proceeding in this way for the other cases, we find the following  complete
list of relations:
$$\eqalignT{ &C_1 C_2 = A_3 D_3,\quad &C_2C_3= A_1D_1\quad & C_3C_1= A_1A_3B_2
\cr &D_1 D_2=B_3C_3^2\quad & D_2D_3= A_1^2B_2B_3\quad & D_1D_3= B_2C_2^2 \cr
&C_1D_1= A_3B_2C_2\quad &C_2D_2= A_1B_3C_3\quad & C_3D_3= A_1B_2C_2\cr}
\eqlabel\zysp$$

The use of the BZ basis to find the elementary couplings and their relations seems
to be novel.\foot{Our construction provides an indirect proof of the validity of
the BZ basis in that from it we recover the result of [\Hongo] derived from the
character method.}

A possible choice of forbidden products is the one given in [\Hongo]: $$\{ C_i
C_j,D_i D_j,C_i D_i\}\eq$$ with $i,j=1,2,3~ {\rm and} ~i\not=j$.
It leads to the generating function:
$$\eqalign{G^{sp(4)}= \left(\prod_{i=1}^3~\tilde{A}_i
\tilde{B}_i\right) &\left( \tilde{C}_1 \tilde{D}_2+D_3 \tilde{C}_1
\tilde{D}_3+C_2 D_1 \tilde{C}_2 \tilde{D}_1 \right.\cr &\left. +C_2 \tilde{C}_2
\tilde{D}_3+D_1
\tilde{C}_3 \tilde{D}_1+C_3 \tilde{C}_3 \tilde{D}_2\right) \cr} \eqlabel\fcr$$
Of course, by modifying the ordering in the Grobner basis, we can get other
choices of forbidden couplings.  Here is another set of forbidden couplings that
can be obtained: $\{D_i D_j, C_i D_i,A_1 D_1,A_3 D_3, A_1 A_3 B_2\}$.
The corresponding generating function reads
$$\eqalign{G^{sp(4)}&=\tilde{B}_1 \tilde{B}_2 \tilde{B}_3
 \left[\left(\prod_{i=1}^3~\tilde{A}_i\right) \tilde{C}_i (1-A_1 A_3 B_2)+D_3
\tilde{D}_3 \tilde{A}_1 \tilde{A}_2 \tilde{C}_1 \tilde{C}_2 \right.\cr
&\quad\left.+D_1 \tilde{D}_1 \tilde{A}_2 \tilde{A}_3 \tilde{C}_2 \tilde{C}_3 
 + D_2 \tilde{D}_2 \tilde{A}_1 \tilde{A}_2 \tilde{A}_3 \tilde{C}_1
\tilde{C}_3 (1-A_1 A_3 B_2) \right] .\cr}
\eq$$
These two generating functions are equivalent when rewritten in terms of the
grading variables, that is, in terms of Dynkin labels.  However, they originate
from two distinct models.  The second one turns out to be well adapted to the
fusion extension.

%================================================================================

\newsec{A vector-basis approach to the construction of generating
functions}

In this section, we present a simple and systematic way
 of generating by hand
all the elementary solutions of a set of linear homogeneous inequalities starting
from the well-known construction of a vector basis.  The first step
amounts to reformulate the system of inequalities in terms of equalities.  We then
look for the elementary independent solutions by relaxing the positivity
requirement.  In  other words, we construct the vector basis.  In a final step,
we find the minimal linear combinations of these vector-basis elements that yield
positive solutions. This will also provide a further illustration of the
MacMahon's projection technique.  But actually, the outcome of this projection is
the desired tensor-product generating function.  Hence, this approach turns out
to be a new way of constructing the generating functions. 
\foot{This method, referred to as being novel,
is probably well-known but we could not trace it precisely in the literature. 
There are implicit remarks in that direction in the first reference of [\Stan].}
This method will
play a key role in our subsequent description of fusion-rule
generating functions.

In the first subsection, we present the equality version of the LR system of
inequalities which holds for $su(N)$. This happens 
to lead to a remarkable graphical description of $su(N)$ tensor products that has
been found by Berenstein and Zelevinsky  [\ref{A.D. Berenstein and A.Z.
Zelevinsky, J. Algebraic Combinat. 1 (1992) 7.}\refname\BZ]. In the following
section, we illustrate the procedure by working out the vector-basis argument for
$su(3)$.  A more general discussion of this technique is presented in the next
section.  We finish with an analysis of the $sp(4)$ case, for which we provide a
novel graphical representation of tensor products.

%------------------------------------------------------------------------
\subsec{$su(N)$ tensor-product basis: from inequalities to equalities: graphical
 representations as BZ triangles for
 $su(N)$}

Consider the direct transformation of
the LR inequalities to equalities by introducing an appropriate number of new
non-negative integer variables.  Consider first the $su(2)$ case, for
which there is a single inequality: $\la_1\geq n_{12}$.  We transform this into
an equality by introducing the positive integer $a$ defined by
$$\la_1= n_{12}+a \eq$$ The expression for $\nu_1$ becomes then
$\nu_1=\la_1+n_{11}-n_{12}=a+n_{11}$.  Since $\mu_1=n_{11}+n_{12}$, we are led
naturally to a triangle representation of the tensor product:
$$\la\otimes \mu\supset \nu \quad \leftrightarrow \quad \matrix{a \cr
n_{12} \quad n_{11} \cr} \quad \eqlabel\wsd$$
We read off the Dynkin label of the $\la$ representation from the sum of the two
integers that form the left side of the triangle, that of the
$\mu$ representation from the bottom of the triangle and the
$\nu_1$ label is the sum of the two integers that form the right side.  A more
uniform notation amounts to setting $a= m_{12}$ and $n_{11}= l_{12}$, in
terms of which the triangle looks quite symmetrical:
$$\matrix{m_{12} \cr
n_{12} \quad\quad l_{12} \cr} \quad \eqlabel\wsdd$$
 with
$$\la_1= m_{12}+n_{12}\qquad \mu_1= n_{12}+l_{12}\qquad \nu_1=m_{12}+l_{12}\eq$$
These numbers $m_{12}$ and $l_{12}$ plays the role of $n_{12}$ in the permuted
versions of the tensor product.
The triangle combinatorial reformulation of the tensor product problem is as
follows: the number of triangles that can be formed from nonnegative
integers $n_{12}, \,m_{12}$ and $l_{12}$ that add up to the Dynkin
labels of the representations under study according to the above relations gives
the multiplicity of the triple coupling $\la\otimes \mu\supset \nu$, or
equivalently, the multiplicity of the scalar representation in the product
$\la\otimes \mu\otimes \nu\supset (0)$ (since for $su(2)$, $\nu^*=\nu$).

For $su(3)$ the situation is somewhat more complicated.  The transformation
of the inequalities (\inee) into equalities takes the form 
$$
\eqalign{ \la_1  &= n_{12}+a\cr
 \la_2 &= n_{13}+b\cr  \la_2+n_{12} &=
n_{13}+n_{23}+c\cr
 n_{11}&= n_{22}+d \cr n_{11}+n_{12}&= n_{22}+n_{23}+e\cr}
\eqlabel\eqee$$
The expression for the other weights becomes
$$\eqalignD{ &\mu_1 = n_{13}+e\qquad
&\mu_2 = n_{22}+n_{23}\cr
&\nu_1 = a+d\qquad
&\nu_2 = n_{22}+c\cr}\eqlabel\others$$
Since there are two expressions for both $n_{11}$ and $\la_2$,
there follows the compatibility relations:
$$n_{12}+d= n_{23}+e\qquad n_{23}+c= b+n_{12}\eq$$
By adding these two  relations, we find:
$$c+d=b+e\eq$$
Again we are led naturally to a triangle representation: with $\zeta= \nu^*$ this
reads
$$\matrix{a_{~}\cr
	n_{12}~~\quad d_{~}\cr
b_{~}~\quad\qquad ~~c_{~}\cr
 n_{13}~\quad e \qquad n_{23} \quad~ n_{22} \cr
}\eqlabel\dtribzz$$
We read the Dynkin labels from the sides of the triangles, from $\la_1$ to
$\zeta_2$ in an anti-clockwise rotation starting from the top of the triangle,
exactly  as for
$su(2)$, except that here there are two labels on each sides.  Notice
that the compatibility conditions amounts to the equality of the sums of
the extremal points of the three pairs of opposite sides of the hexagon obtained
by dropping the three corners of the triangle.

 Again a more symmetrical notation
calls for itself:
$$a= m_{13}\qquad b= m_{23}\qquad  c= m_{12}\qquad  d= l_{23}\qquad  e= l_{12}\qquad 
 n_{22}= l_{13}\qquad\eq$$
in terms of which the triangle reads
$$\matrix{m_{13}\cr
	n_{12}~~\quad l_{23}\cr
m_{23}~\quad\qquad ~~m_{12}\cr
 n_{13}~\quad l_{12} \qquad n_{23} \quad~ l_{13} \cr
}\eqlabel\trbz$$ 
with labels fixed by: 
$$\eqalignD{&\la_1=m_{13}+n_{12}\quad &\la_2= m_{23}+n_{13}\cr
&\mu_1= n_{13}+l_{12} \quad &\mu_2= n_{23}+l_{13}\cr
&\zeta_1= l_{13}+m_{12}\quad &\zeta_2= l_{23}+m_{13}\cr}
\eqlabel\wee$$
The hexagon conditions read:
$$\eqalign{n_{12}+m_{23}& = n_{23}+m_{12},\cr
l_{12}+m_{23} &= l_{23}+m_{12},\cr 
l_{12}+n_{23} &= l_{23}+n_{12}.\cr
}\eqlabel\dtribz$$
In terms of triangles, the problem of finding the multiplicity of the $su(3)$
tensor product $\la\otimes \mu\otimes \zeta\supset 0$ boils down to enumerating
the number of triangles made with nonnegative integers that form a bipartition of
the Dynkin labels and that  satisfy the above three  hexagon relations.

For $su(4)$ the BZ triangle is obtained in a similar way.  One first transforms
the inequalities (\inequatre) into equalities by introducing positive integers $a,
b, \cdots,
\kappa$:
$$\eqalign{
\la_1&=  n_{12}+a\cr
\la_2& = n_{13}+b\cr
\la_2+n_{12}& = n_{13}+n_{23}+c\cr
\la_3& = n_{14}+d\cr
\la_3+n_{13}& = n_{14}+n_{24}+e\cr
\la_3+n_{13}+n_{23}& = n_{14}+n_{24}+  n_{34}+f\cr
n_{11}& = n_{22}+g\cr
n_{11}+n_{12}& = n_{22}+ n_{23}+h\cr
n_{11}+n_{12}+n_{13}& = n_{22}+ n_{23}+n_{24}+i\cr
n_{22}& = n_{33}+j\cr
n_{22}+n_{23}& = n_{33}+ n_{34}+\kappa\cr
}\eqlabel\equatre$$ 
The Dynkin labels of $\mu$ and $\zeta=\nu^*$ becomes
$$\eqalignT{ &\mu_1 = n_{14}+i, \quad &\mu_2= n_{24}+\kappa, \quad &
\mu_3=
n_{33}+n_{34}\cr
&\zeta_1 = n_{33}+f, \quad &\zeta_2= c+j, \quad &\zeta_3=
a+g\cr}\eq$$
Since every Dynkin label is the sum of two positive integers, a triangle
representation is again natural:
$$\matrix{a_{~}\cr
	n_{12}~~\quad g_{~}\cr
x_{~}~\qquad\qquad ~~u_{~}\cr
y_{~}\qquad ~~\qquad ~~ \qquad v_{~}\cr
d_{~}\qquad\qquad\quad ~~\qquad\qquad\quad f_{~} \cr
n_{14}\qquad~~ i_{~}\qquad~ s_{~}\quad~\quad t_{~}\quad~~~ n_{34}\qquad
n_{33} \cr  }\eqlabel\fobzz$$
However the position of the integers specifying the labels 2 is ambiguous at
this point: $x$ can be either $b$ or $n_{13}$ ($y$ being the other one),
similarly
$(u,v)$ is related to the doublet
$(j,c)$ and
$(s,t)$ to
$(n_{24},\kappa)$.  Moreover not all the needed integers appear in this triangle:
$e,h,n_{23}$ are missing. In order to take into account the various
compatibility relations, it is natural to insert the three remaining points in
the center of the big triangle forming then three hexagons:
$$\matrix{
\,&\,&\,&\,&\,&\bullet&\,&\,&\,&\,&\,\cr
\,&\,&\,&\,&\bullet&\,&\bullet&\,&\,&\,&\,\cr
\,&\,&\,&\bullet&\,&\,&\,&\bullet&\,&\,&\,\cr
\,&\,&\bullet&\,&\bullet&\,&\bullet&\,&\bullet&\,&\,\cr
\,&\bullet&\,&\,&\,&\bullet&\,&\,&\,&\bullet&\,\cr
\bullet&\,&\bullet&\,&\bullet&\,&\bullet&\,&\bullet&\,&\bullet\cr}\eq$$
Quite remarkably,
the different conditions are simply the equality of the opposite sides of every
hexagon, exactly as for $su(3)$, and these completely fix the position of every
integers in the triangle.  For instance, replacing $\la_2=n_{13}+b$ into
$\la_2+n_{12}=n_{13}+n_{23}+c$ yields
$$n_{12}+b=n_{23}+c\eq$$
For this to be an hexagon relation,  $b$ and $c$ must belong to the same hexagon
and since they are on opposite sides of the triangles, they must belong to the
first hexagon: hence $x=b$ and $u=c$.  That also fixes the position of $n_{23}$
inside the big triangle.  Proceeding in this way with the other constraints, we
end up with the following representation:
$$\matrix{a_{~}\cr
	n_{12}~~\quad g_{~}\cr
b_{~}~\qquad\qquad ~~c_{~}\cr
n_{13}\qquad h_{~}\qquad n_{23} \qquad j_{~}\cr
d_{~}\qquad\qquad\quad e_{~}\qquad\qquad\quad f_{~} \cr
n_{14}\qquad~ i_{~}\qquad~ n_{24}\quad~\quad \kappa_{~}\quad~~
 n_{34}\qquad
n_{33} \cr  }\eqlabel\fobzzz$$

Finally, a more symmetrical form is obtained by redefining the name of the
positive integers $a,\cdots,\kappa$ and $n_{33}$  as follows:
$$\matrix{m_{14}\cr
	n_{12}~~\quad l_{34}\cr
m_{24}~\qquad\qquad ~~m_{13}\cr
n_{13}\qquad l_{23}\qquad n_{23} \qquad l_{24}\cr
m_{34}\qquad\qquad\quad m_{23}\qquad\qquad\quad m_{12} \cr
n_{14}\qquad~~ l_{12}\qquad~ n_{24}\quad~\quad l_{13}\quad~~~ n_{34}\qquad
l_{14} \cr  }\eqlabel\fobz$$
They are related to the Dynkin labels by
$$\eqalignT{ &\lambda_1=m_{14}+n_{12}\quad &\lambda_2=m_{24}+n_{13}\quad
& \lambda_3 =m_{34}+n_{14} \cr 
&\mu_1= n_{14}+l_{12} \quad
&\mu_2= n_{24}+l_{13} \quad &\mu_3=  n_{34}+l_{14} \cr
&\zeta_1= l_{14}+m_{12}
\quad &\zeta_2=
l_{24}+m_{13}\quad &\zeta_3= l_{34}+m_{14}
\cr }\eqlabel\mma$$
and the  hexagon relations read:
$$\eqalignT{ & n_{12}+m_{24} =m_{13}+n_{23}\ \quad &n_{12}+l_{34}
=l_{23}+n_{23}\quad &m_{24}+l_{23} =l_{34}+m_{13}\ \cr
&n_{13}+l_{23} =l_{12}+n_{24}\quad &n_{13}+m_{34} =n_{24}+m_{23}\quad
&m_{34}+l_{12} = l_{23}+m_{23}\cr
&l_{24}+n_{23} =l_{13}+n_{34}\quad\ &n_{23}+m_{23} =m_{12}+n_{34}
\quad\ &l_{13}+m_{23} =l_{24}+m_{12} \cr }\eqlabel\mmb$$

The $su(N)$ generalisation is obvious; the triangle is built out of
$(N-1)(N-2)/2$ hexagons and three corner points.

Here is the rationale for the labelling $n_{ij},m_{ij},l_{ij}$ from the triangle
point of view [\BZ]. If
$e_i$ are orthonormal vectors in ${\bf R}^N,$ then the positive roots of $su(N)$
can be represented in the form $e_i-e_j,\ 1\leq i<j\leq N.$ The triangle encodes
three sums of positive roots:
$$\eqalign{\mu +\zeta -\lambda^*\ &=\ \sum_{i<j}\ l_{ij} (e_i-e_j)\ \ ,\cr
\zeta +\lambda -\mu^*\ &=\ \sum_{i<j}\ m_{ij} (e_i-e_j)\ \ ,\cr
\lambda +\mu -\zeta^*\ &=\ \sum_{i<j}\ n_{ij} (e_i-e_j)\ \ ,\cr}
\eqlabel\expansions$$
 The hexagon relations are simply the consistency
conditions for these three expansions. Clearly, the variables $n_{ij}$ that
appear in the above relations are exactly the $n_{ij}$ that appear in the LR
tableaux for the product $\la\otimes \mu\supset\zeta^*=\nu$.

%========================================================================

\subsec{From a vector basis to the generating function: the $su(3)$ case}

Given the transcription of inequalities into equalities, we can easily extract
the corresponding basis vectors.  This is the starting point of a new method
for constructing the tensor-product generating functions.
%à cut: (which is described in the first reference in [\Stan])
To keep things
concrete, we focus on the $su(3)$ case.  The goal is to first get a vector basis
and then to project it to get the elementary couplings.  The generating function
is a direct result  of this procedure.

The equality version of 
the LR inequalities have already been presented in the previous subsection:
these are (\wee) and (\dtribz); they underlie the construction of the BZ triangle
(\trbz). 
The last hexagon condition of (\dtribz) is the difference of the previous two so
it is not an independent relations. We thus have a total of 15 variables: $\la_1,
\cdots,
\nu_2, l_{12}, \cdots , n_{23}$ and 8 equations. The number of independent
variables is thus 7. These will be chosen to be
$m_{13},m_{23},l_{13},l_{23},n_{12},n_{13},n_{23}$. The dependent variables are
fixed as follows:
$$\eqalign{ \lambda_1&= m_{13}+n_{12} \cr
							     \lambda_2&= m_{23}+n_{13}  \cr
					       \mu_1 &= n_{13}+n_{12}+l_{23}-n_{23}				\cr
            \mu_2 &= n_{23}+l_{13}				\cr
            \zeta_1 &= n_{12}+m_{23}+l_{13}-n_{23}				\cr
												\zeta_2 &= l_{23}+m_{13}				\cr
            l_{12} &= n_{12}+l_{23}-n_{23}				\cr
            m_{12} &= n_{12}+m_{23}-n_{23}				\cr} \eq $$
We now look for the elementary solutions of this system (without invoking
the constraint that all the above dependent variables should be necessarily
positive).  The sought basis vectors are obtained by setting one of the
variable 
$m_{13},\cdots,n_{23}$ to 1 and all other set equal to zero.  This produces (in
order) the triangles $E_2, E_5, E_6, E_3, E_7, E_4$ and $Z_1$ displayed below:
$$\matrix{E_2: (1,0)(0,0)(0,1)\cr~\cr\tri000100000}\qquad
\matrix{E_3: (0,0)(1,0)(0,1)\cr~\cr\tri000010001}$$
$$\matrix{E_4: (0,1)(1,0)(0,0)\cr~\cr\tri100000000}\qquad
\matrix{E_5: (0,1)(0,0)(1,0)\cr~\cr\tri010001000}\qquad
\matrix{E_6: (0,0)(0,1)(1,0)\cr~\cr\tri000000100}$$
$$\matrix{E_7: (1,0)(1,0)(1,0)\cr~\cr\tri001001001}\qquad
\matrix{Z_1: (0,0)(-1,1)(-1,0)\cr~\cr\tri00000{-1}01{-1} } \eqlabel\zz$$
These are all genuine BZ triangles except for $Z_1$ which has some negative
entries.  However, at this level, there are no relations
between these elementary solutions (the basis vectors are independent), hence
the decomposition of any solution in terms of these 7 basic ones is unique.  All
solutions are then freely generated from the following function:
$$G={1 \over (1-E_2) (1-E_3) (1-E_4) (1-E_5) (1-E_6) (1-E_7) (1-Z_1)}\eq$$
To recover the generating function for all tensor products from the above
expression, we need to project out terms that lead to triangles with negative
entries.  To achieve this, we introduce the grading variables associated to the
above couplings (compare the above triangles with the general form given in
(\trbz)):
$$\eqalignT{ &E_2: M_{13}, ~\qquad &E_3: L_{12}L_{23}\qquad &E_4: N_{13}\cr
 &E_5: M_{12}M_{23}\qquad &E_6: L_{13}\qquad &E_7: L_{12}M_{12}N_{12}\cr
& ~ &Z_1:L_{12}^{-1}M_{12}^{-1}N_{23}  &\cr}\eq$$
Our generating function follows from the projection of the above function $G$, 
re-expressed in terms of the grading variables, to positive powers of $L_{12}$ and
$M_{12}$.  Equivalently, one can rescale $L_{12}$ by $x$ and $M_{12}$ by $y$ and
project to positive powers of $x$ and $y$ and set $x=y=1$ in the result.  This is
equivalent to the rescaling
$$E_3\rw xE_3\qquad E_5\rw yE_5\qquad E_7\rw xyE_7\qquad Z_1\rw
x^{-1}y^{-1}Z_1\eq$$
We are thus led to consider 
$$\Ox\Oy \; G(E_2, xE_3, \cdots, x^{-1}y^{-1}Z_1)\eq$$
Keeping only those terms which depend explicitly upon $x$ or $y$, we have then
$$\eqalign{\Ox\Oy \; &{1 \over  (1-xE_3)(1-yE_5)  (1-xyE_7)
(1-x^{-1}y^{-1}Z_1)}\cr
& \quad= {1 \over  (1-xE_3)(1-yE_5)  (1-E_7Z_1)}\left({1\over
1-xyE_7}+{x^{-1}y^{-1}Z_1\over 1-x^{-1}y^{-1}Z_1}\right)\cr}\eq$$
No more work is need for the first term. For the second one, we have
$$\eqalign{\Ox\Oy \; &{x^{-1}y^{-1}Z_1 \over  (1-xE_3) (1-E_7Z_1)(1-x^{-1}Z_1E_5)
}\left({yE_5\over 1-yE_5}+{1\over 1-x^{-1}y^{-1}Z_1}\right)\cr
&=\Ox\; {x^{-1}E_5Z_1 \over (1-E_5)(1-E_7Z_1) (1-xE_3) (1-x^{-1}Z_1E_5)
}\cr
&=\Ox\; {x^{-1}E_5Z_1 \over (1-E_5)(1-E_7Z_1) (1-E_3E_5Z_1)}\left({xE_3\over 
1-xE_3}+{1\over 1-x^{-1}Z_1E_5 }\right)\cr
&= {E_3E_5Z_1\over (1-E_5)(1-E_7Z_1) (1-E_3E_5Z_1)(1-E_3)}\cr}\eq$$
We then introduce  the following two new elementary couplings
$$E_1=E_7 Z_1  \quad \quad E_8= E_3 E_5 Z_1\eq$$
Collecting the two terms resulting from the projection, we end up with
$$G^{su(3)}= \left(\prod_{i=1}^8 \tilde{E}_i\right) (1-E_7E_8)\eq$$
which is indeed the $su(3)$ tensor-product generating function.

As a side remark, we indicate how the various triangles associated to a given
triple product are related to each others.  For this question, we consider the
three weights to be fixed.  The system of equations is now nonhomogeneous and
solutions are given by a linear combination of a particular solution and the sum
of all homogeneous solutions, that is, solutions of the system with all Dynkin
labels set equal to zero:
$$\eqalign{ 0= & m_{13}+n_{12} \cr
							     0= & m_{23}+n_{13}  \cr
					       0 = & n_{13}+n_{12}+l_{23}-n_{23}				\cr
            0 = & n_{23}+l_{13}				\cr
            0 = & n_{12}+m_{23}+l_{13}-n_{23}				\cr
												0 = & l_{23}+m_{13}				\cr
            0 = & n_{12}+l_{23}-n_{23}-l_{12}				\cr
            0 = & n_{12}+m_{23}-n_{23}-m_{12}				\cr} \eq $$
These solutions are given by
$n ~\Delta$ with $n\in \Z$
and
$$\Delta=\tri{~~1}{-1~~}{-1}{1}{-1}{-1}{1}{-1~}{-1~}
\eq$$ Hence, given an allowed triangle, all other triangles related to the same
tensor product can be obtained by adding or subtracting a number of times
$\Delta$ while ensuring that all triangle entries are positive [\ref{L. B\'egin, P. Mathieu
and M.A. Walton, Mod. Phys. Lett. {\bf A7}
 (1992) 3255.}\refname\LB].  

The great advantage of the vector-basis procedure is that the basis vectors are
very easily obtained and their number grows slowly with $N$ for $su(N)$ (their
number being
$(N-1) (N+4)/2$) 
as compared to the number of tensor-product elementary couplings:
$$\eqalignT{ &{\rm algebra}  &\quad  \# ~{\rm elem.~coupl.}   &\quad {\rm dim.~
vect.~space} 
\cr
           &su(2)& \qquad 3  &\qquad3 \cr
           &su(3)  & \qquad8  &\qquad 7 \cr
           &su(4)  & \qquad 18  &\qquad  12 \cr
           &su(5)  &\qquad  45  & \qquad 18 \cr
           &su(6)  & \qquad 138  &\qquad  25 \cr
           &su(7)& \qquad 526  & \qquad 33 \cr } \eq$$

%==============================================================================
\subsec{General aspects of the vector-basis construction}

In general, of
course, the fundamental solutions to the linear system may have non-integral
values of the variables. However the corresponding terms
in the generating function can be eliminated by
rationalising all the denominator terms and then
keeping only those terms in the numerator that
have integral exponents. This suggests the following
modification of MacMahon's algorithm.

Consider the system of equations
$$Mx=0,\quad x\in\N^k\eqlabel\Meq$$
where $M$ is a matrix of rank $s$.  We thus have $k$ variables and $s$ relations
between them. The dimension of the vector basis is thus $k-s$.  We will denote
the independent (free) variables as $x_i$, $i=1, \cdots, k-s$ and the remaining
ones as ${\tilde x}_j$, $j=1, \cdots, s$.  To find a generating function for the
solutions of this system:

\item{\bf 1.} First construct a basis in $\Q^k$ for the solutions
of $Mx=0$ 
by setting $x_i=1$ with all other $x_j$ zero ($j=1, \cdots, k-s,
\, j\not=i$).  Denote by
${\tilde x}_j^{(1)}$ the value of the dependent variable ${\tilde x}_j$
evaluated at
$x_1=1$ with all other $x_i$ zero. The basis then reads
$$
\eqalign{ &\epsilon_1=(1,0,0\dots, 0; \{{\tilde x}_j^{(1)}\}),\cr
 &\epsilon_2=(0,1,0\dots, 0; \{{\tilde x}_j^{(2)}\}),\cr
 &\cdots\cr
&\epsilon_{k-s}=(0,0,0,\dots,1; \{{\tilde x}_j^{(k-s)}\})\cr}\eq
$$
By construction, the $\epsilon_i$'s are linearly independent.  
However notice that in general the ${\tilde x}_j^{(i)}$ might be rational.

\item{\bf 2.} From the form of the $\epsilon_i$'s, it follows that
any solution to (\Meq) can be written as $\sum_i c_i\epsilon_i$
with $c_i$ non-negative integers. In particular this means
that every solution to (\Meq) corresponds to
a term in the generating function:
$$G(X)={1\over{(1-X^{\epsilon_1})(1-X^{\epsilon_2})\dots(1-X^{\epsilon_s})}}\eq$$
where $X_1,\dots,X_k$ are grading variables.

\item{\bf 3.} $G(x)$ may contain negative or fractional
exponents due to the occurrence of ${\tilde x}_j^{(i)}$ in the exponents. These are
eliminated by first using MacMahon's algorithm to eliminate any negative
exponents and then rationalizing denominators and keeping only terms
with integral exponents in the numerators.

The result is the generating function for the solutions
to (\Meq). This algorithm, however, does not seem to be
optimal in all case. Consider the following example:
$$\eqalign{
3x-2y+z-3t&= 0\cr \quad 2x+y-2z-t& =0\cr}\eqlabel\exeq$$
with $ x,y,z,t\in\N$.
Using linear algebra, we find two basic solutions:
$({3\over 7},{8\over 7},1,0)$ and
$({5\over 7},-{3\over 7},0,1)$. This means that
the initial generating function is
$$
{1\over{(1-X^{3/7}Y^{8/7}Z)(1-X^{5/7}Y^{-3/7}T)}}
$$
from which we must eliminate negative and
fractional exponents. This is somewhat lengthy.
However, the calculation can be shortened by some
observations. First note that we can take
as the fundamental solutions:
$ ({3\over {56}},{1\over 7},{1\over 8},0)$ and
$({5\over {21}},-{1\over 7},0,{1\over 3})$ since
once again any solution to (\exeq) is a non-negative
linear combination of these two solutions.
This give as the initial generating function:
$$
{1\over{(1-X^{3/56}Y^{1/7}Z^{1/8})(1-X^{5/{21}}Y^{-1/7}T^{1/3})}}
\eq$$
Keeping the positive exponents in $Y$ yields:
$$
{1\over{(1-X^{7/24}Z^{1/8}T^{1/3})(1-X^{3/56}Y^{1/7}Z^{1/8})}}
\eq$$
and keeping integral exponents in $T$ and $Y$ gives:
$$
{1\over{(1-X^{7/8}Z^{3/8}T)(1-X^{3/8}YZ^{7/8})}}.
\eq$$ 
Rationalizing these denominators and keeping 
integral exponents in the numerator yields the
final generating function:
$$
\eqalign{
( 1+{X}^{2}{Z}^{3}T{Y}^{3}+{X}^{4}{Z}^{6}{T}^{2}{Y}^{6}+{X}^{3}{Z}
^{2}{T}^{3}Y+&{X}^{5}{Z}^{5}{T}^{4}{Y}^{4}+\cr
{X}^{7}{Z}^{8}{T}^{5}{Y}^{7}+{X}^{6}{Z}^{4}{T}^{6}{Y}^{2}+{X}^{8}{Z}^{7}{T}^{7}{Y}^{5})&
(1-{X}^{7}{Z}^{3}{T}^{8})^{-1} (1-{X}^{3}{Y}^{8}{Z}^{7} )^{-1}\cr}
\eq$$

Notice however that in applications to tensor products, it seems we never encounter
rational exponents but simply negative ones.  The procedure is thus usually
simpler.
%==================================================================

\subsec{Multiple $su(2)$ products from the vector-basis construction}

A simple and different application of the formalism just developed is furnished
by the analysis of $su(2)$ quadruple tensor products.  This application is
different in that it does not rely on the triangle description and as such, its
formulation is less direct.
\foot{This does not mean however that there are no diagrammatic representations
for the quadruple product. 
In fact, having a set of inequalities, we can transform then into equalities, as
it is done below, and from them set up a diagrammatic representation. In the
present case, it could correspond to two adjacent
$su(2)$ triangles, one upside down, with their adjacent sides forced to be
equal.  Here we simply mean that the analysis will not rely on such
a description.}
 It will serve as a preparation the somewhat more complicated
$sp(4)$ example treated in the following section. 

The Diophantine description of this problem has been presented in section 4.2. It
is based on the two inequalities (\multiin) which are readily transformed into
equalities by the introduction of two positive integers $a_1, \, a_2$:
$$\la_1= n_{12}+a_1\qquad \la_1+n_{11}-n_{12}= m_{12}+a_2\eq$$
However this system does not contain any reference to the variable $m_{11}$ and
for this reason we introduce the further constraint $m_{11}\geq 0$ which calls
for a new integer variable:
$$m_{11}= a_3\eq$$
We have thus a total of 8 variables : $\{\la_1, 
n_{11}, n_{12},m_{11},m_{12}, a_1, a_2, a_3\}$ and 3 equations.  There are thus 5
independent variables, chosen to be  $\{a_1, a_2, a_3, n_{12}, m_{12}\}$.  The
basis vectors, with components ordered as follows $$(a_1, a_2,
a_3, n_{12}, m_{12}; \la_1, n_{11}, m_{11})\eq$$ are obtained by successively
setting equal to 1 one of $\{a_1, a_2, a_3, n_{12}, m_{12}\}$ and the others
equal to 0.  These basis vectors together with their exponentiated version written
in terms of appropriate grading variables read:
$$\eqalignT{
& (1,0,0,0,0;1,-1,0)\quad &: L_1N_{11}^{-1}A_1\cr
& (0,1,0,0,0;0,1,0)\quad &: N_{11}A_2\cr
& (0,0,1,0,0;0,0,1)\quad &: M_{11}A_3\cr
& (0,0,0,1,0;1,0,0)\quad &: L_1N_{12}\cr
& (0,0,0,0,1;0,1,0)\quad &: N_{11}M_{12}\cr
}\eq$$
The desired generating function is obtained from the projection to positive powers
of
$N_{11}$ of the function
$${1\over (1-L_1N_{11}^{-1}A_1)(1- N_{11}A_2)(1-L_1N_{12}) (1-
N_{11}M_{12}) (1- M_{11}A_3) }\eq$$
The projection operation is done by the familiar method and the result, after
setting all $A_i=1$ is
$$ G= {1-L_1N_{11}M_{12}\over  (1- L_1N_{12})(1-
L_1M_{12})(1-L_1) (1-N_{11}M_{12})(1- N_{11} )(1- M_{11})}\eq$$ from which we
read of the 6 elementary couplings $E_1, \cdots , E_6$ (in the order where they
appear in the denominator) given in (\mutipo) and the relation
$E_3E_4= E_2E_5$.  The above function is exactly the one derived in section 4.2.

%------------------------------------------------------------------------

\subsec{$sp(4)$
diamonds and the vector-basis derivation of the generating function}

The system of inequalities (\bzspq ) pertaining to $sp(4)$ can be transformed
into a system of equations in the standard way: by setting 
$r_1/2={\cal R}_1$ and $r_2/2={\cal R}_2$ and 
introducing the integers $a_i$, we get:
\foot{The original idea of looking for a diagrammatic representation of $sp(4)$
tensor products along theses lines is due to M. Walton.} 
$$
\eqalign{\lambda_1 & =p_{12}+a_1 \cr
									\lambda_2 & ={\cal R}_1+a_2 \cr
\mu_1 & =q_{12}+a_5  \cr 
									\mu_2 & ={\cal R}_2+a_8 \cr
\nu_1 & =a_1+a_7 \cr
									\nu_2 & =a_4+a_8 \cr
a_2+p_{12} &
=a_3+q_{12} \cr
									a_3+{\cal R}_1 & =a_4+{\cal R}_2 \cr
a_5+2{\cal R}_{2} & =a_6+2{\cal R}_{1} \cr
									a_6+q_{12} & =a_7+p_{12} \cr}   \eq$$
This leads to a diamond-type graphical representation of the tensor product that
has the advantage over the one presented in [\BZ] of being linear (the sum of two
diamonds is also a diamond):

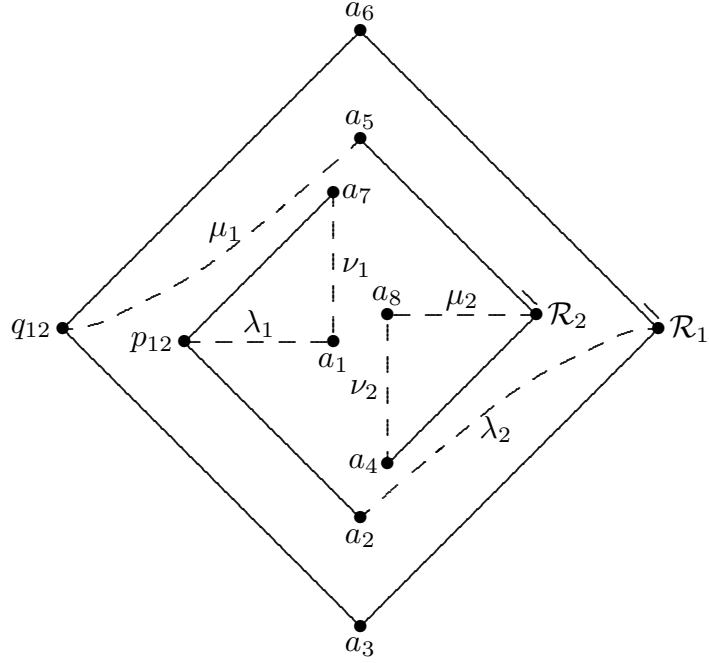
\input Pictex

\def\bpc{\beginpicture}

\def\epc{\endpicture}

\def\figureF{
\bpc
	\setcoordinatesystem units <0.6mm,0.6mm>
	\setplotarea x from -55 to 55, y from -55 to 55
	\setlinear
	\plot 0 55 -55 0 0 -55 55 0 0 55 /
	\plot 5 -25 32.5 2.5  0 35 /
\plot -5 25 -32.5 -2.5  0 -35 /
\plot 32.5 4.5 30 7 /
	\plot 55 2 52.5 4.5 /
	\multiput {$\bullet$} at 0 55 0 -55 55 0 -55 0 0 35 0 -35 32.5 2.5 -32.5 -2.5 5 -25 -5 25 5 2.5 -5 -2.5
/
	\put {${\cal R}_1$} [l] at 	57 0
	\put {${\cal R}_2$} [l] at 34.5 2.5
	\put {$q_{12}$} [r] at -57 0
	\put {$p_{12}$} [r] at -34.5 -2.5
	\put {$\lambda_1$} [b] at -18.75 -1.5
	\put {$\mu_2$} [b] at 18.75 3.5
	\put {$\nu_1$} [l] at -3.5 11.25
	\put {$\nu_2$} [r] at 3.5 -11.25
	\put {$\mu_1$} [b] at -25 16
	\put {$\lambda_2$} [t] at 25 -16
	\put {$a_6$} [b] at 0 57
	\put {$a_5$} [b] at 0 37
	\put {$a_2$} [t] at 0 -37
	\put {$a_3$} [t] at 0 -57
	\put {$a_4$} [r] at 3.5 -25
	\put {$a_7$} [l] at -3.5 25
	\put {$a_8$} [b] at 5 4.5
	\put {$a_1$} [t] at -5 -4.5
	\setdashes
	\plot 32.5 2.5 5 2.5 5 -25 /
	\plot -32.5 -2.5 -5 -2.5 -5 25 /
	\setquadratic
	\plot 0 -35 5 -30.5 15 -22 20 -18 25 -14 30 -10.5 35 -7.5 40 -5 45 -2.5 50 -0.5 55 0
/
	\plot 0 35 -5 30.5 -15 22 -20 18 -25 14 -30 10.5 -35 7.5 -40 5 -45 2.5 -50 0.5 -55 0
/
\epc}

\vskip1cm \centerline{ \figureF} \vskip1cm

In this picture, all data pertaining to the first (second) Dynkin label appear at
the left (right).
Dotted lines relate those two points that compose the label indicated beside it. 
Opposite continuous lines are constrained to be equal, with the length of a line
being defined as the sum of its extremal points except for the lines delimited by
the points
$(a_6, {\cal R}_1)$ and $(a_5,{\cal R}_2)$ where the point ${\cal R}_i$ is
counted twice (the little bar besides ${\cal R}_1$ and ${\cal R}_2$ being a
reminder of this particularity).  Explicitly, for those lines, we have thus the
constraint
$a_6+2{\cal R}_1=a_5+2{\cal R}_2$.  Given a triple
$sp(4)$ product,  the number of such diamonds that can be drawn with non-negative
entries yields the multiplicity of the product.  
For instance, the two diamonds that describe the triple coupling $(1,1)\otimes
(1,1) \otimes (2,0)$ are:

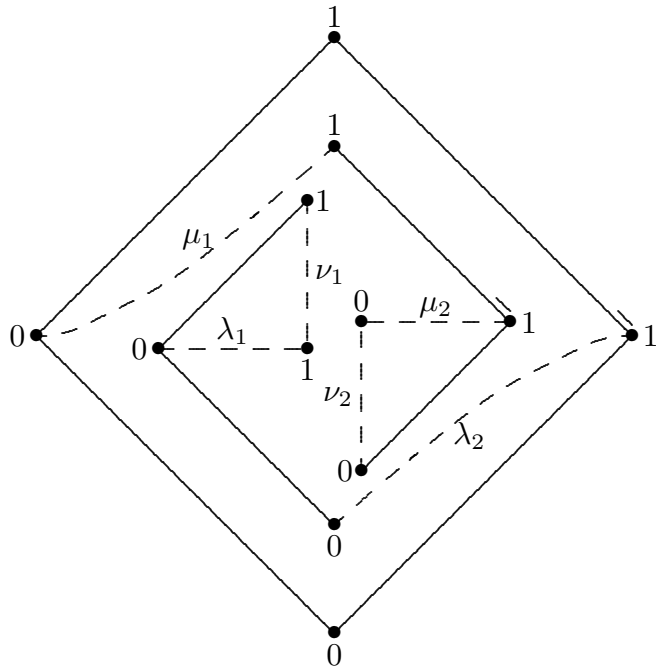
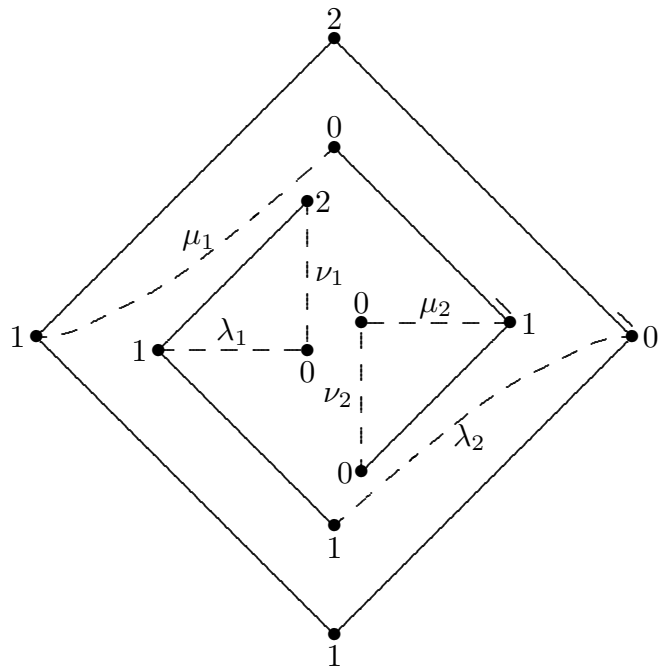
\def\figureFpex{
\bpc
	\setcoordinatesystem units <0.6mm,0.6mm>
	\setplotarea x from -55 to 55, y from -55 to 55
	\setlinear
	\plot 0 55 -55 0 0 -55 55 0 0 55 /
	\plot 5 -25 32.5 2.5  0 35 /
	\plot -5 25 -32.5 -2.5  0 -35 /
	\plot 32.5 4.5 30 7 /
	\plot 55 2 52.5 4.5 /
	\multiput {$\bullet$} at 0 55 0 -55 55 0 -55 0 0 35 0 -35 32.5 2.5 -32.5 -2.5 5 -25 -5 25 5 2.5 -5 -2.5
/
	\put {1} [l] at 	57 0
	\put {1} [l] at 34.5 2.5
	\put {0} [r] at -57 0
	\put {0} [r] at -34.5 -2.5
	\put {$\lambda_1$} [b] at -18.75 -1.5
	\put {$\mu_2$} [b] at 18.75 3.5
	\put {$\nu_1$} [l] at -3.5 11.25
	\put {$\nu_2$} [r] at 3.5 -11.25
	\put {$\mu_1$} [b] at -25 16
	\put {$\lambda_2$} [t] at 25 -16
	\put {1} [b] at 0 57
	\put {1} [b] at 0 37
	\put {0} [t] at 0 -37
	\put {0} [t] at 0 -57
	\put {0} [r] at 3.5 -25
	\put {1} [l] at -3.5 25
	\put {0} [b] at 5 4.5
	\put {1} [t] at -5 -4.5
	\setdashes
	\plot 32.5 2.5 5 2.5 5 -25 /
	\plot -32.5 -2.5 -5 -2.5 -5 25 /
	\setquadratic
	\plot 0 -35 5 -30.5 15 -22 20 -18 25 -14 30 -10.5 35 -7.5 40 -5 45 -2.5 50 -0.5 55 0
/
	\plot 0 35 -5 30.5 -15 22 -20 18 -25 14 -30 10.5 -35 7.5 -40 5 -45 2.5 -50 0.5 -55 0
/
\epc}

\vskip1cm \centerline{\figureFpex} \vskip1cm

\def\figureFdex{
\bpc
	\setcoordinatesystem units <0.6mm,0.6mm>
	\setplotarea x from -55 to 55, y from -55 to 55
	\setlinear
	\plot 0 55 -55 0 0 -55 55 0 0 55 /
	\plot 5 -25 32.5 2.5  0 35 /
	\plot -5 25 -32.5 -2.5  0 -35 /
	\plot 32.5 4.5 30 7 /
	\plot 55 2 52.5 4.5 /
	\multiput {$\bullet$} at 0 55 0 -55 55 0 -55 0 0 35 0 -35 32.5 2.5 -32.5 -2.5 5 -25 -5 25 5 2.5 -5 -2.5
/
	\put {0} [l] at 	57 0
	\put {1} [l] at 34.5 2.5
	\put {1} [r] at -57 0
	\put {1} [r] at -34.5 -2.5
	\put {$\lambda_1$} [b] at -18.75 -1.5
	\put {$\mu_2$} [b] at 18.75 3.5
	\put {$\nu_1$} [l] at -3.5 11.25
	\put {$\nu_2$} [r] at 3.5 -11.25
	\put {$\mu_1$} [b] at -25 16
	\put {$\lambda_2$} [t] at 25 -16
	\put {2} [b] at 0 57
	\put {0} [b] at 0 37
	\put {1} [t] at 0 -37
	\put {1} [t] at 0 -57
	\put {0} [r] at 3.5 -25
	\put {2} [l] at -3.5 25
	\put {0} [b] at 5 4.5
	\put {0} [t] at -5 -4.5
	\setdashes
	\plot 32.5 2.5 5 2.5 5 -25 /
	\plot -32.5 -2.5 -5 -2.5 -5 25 /
	\setquadratic
	\plot 0 -35 5 -30.5 15 -22 20 -18 25 -14 30 -10.5 35 -7.5 40 -5 45 -2.5 50 -0.5 55 0
/
	\plot 0 35 -5 30.5 -15 22 -20 18 -25 14 -30 10.5 -35 7.5 -40 5 -45 2.5 -50 0.5 -55 0
/
\epc}

\centerline{\figureFdex}\vskip1cm

The dimension of the vector basis is 8 (18 variables and 10 equations, the
last four equations above being linearly independent).  As our free variables, we
choose the set $\{\Rc_1, \Rc_2, p_{12}, q_{12}, a_1, a_3, a_6, a_8\}$.  The 8
basis vectors in terms of grading variables are
$$\eqalignD{
& E_1: L_2M_1^2N_2A_4A_5^2R_1\qquad &E_2:
M_1^{-2}M_2N_2^{-1}A_4^{-1}A_5^{-2}R_2\cr & E_3:
L_1L_2^{-1}N_1^{-1}A_2^{-1}A_7^{-1}P_{12}\qquad &E_4: L_2M_1N_1A_2A_7Q_{12}\cr
&E_5: L_1N_1A_1\qquad &E_6: L_2N_2A_2A_3A_4\cr
& E_7: M_1N_1A_5A_6A_7\qquad &E_8 : M_2N_2A_8\cr}\eq$$ 
The generating function is obtained by first projecting of the function
$\prod(1-E_i)^{-1}$ to positive powers for each grading variables and then by
setting all grading variables equal to 1 except for $L_i, M_i, N_i$'s.  The
$sp(4)$ elementary couplings  elementary couplings are simple products of the
$E_i$'s (the following
$A_{1,2,3}$ should not be confused with the above grading variables):
$$\eqalignT{
&A_1=E_7\qquad &A_2= E_5\qquad & A_3= E_3E_4 \cr 
&B_1=E_8\qquad &B_2= E_6\qquad & B_3= E_1E_2 \cr
&C_1=E_4\qquad &C_2= E_2E_3E_6E_7^2\qquad & C_3= E_1E_3E_7 \cr
&D_1=E_2E_3^2E_6^2E_7^2\qquad &D_2= E_1\qquad & D_3= E_2E_6E_7^2
\cr}\eq$$
The complete list of $sp(4)$ elementary couplings (\spele) are thus recovered.

%=========================================================================

\newsec{Summary and conclusion}

\subsec{Listing the methods considered}

Here is the list of methods that have been reviewed here for construction
tensor-product generating functions:

\n 1- {\it The character method}

This has been extensively exemplified in section 2.  It starts from first
principles (the multiplication of two characters or more precisely, two character
generating functions) but it is computationally complicated.

\n 2- {\it Composition of generating function}

This technique uses other, simpler generating functions to construct the
required generating function. The method has been illustrated in
sections 2.3 and 2.5.  One example of this method is the  use of
the Giambelli formula.

\n 3- {\it Diophantine inequalities  -- Elementary couplings and
relations}

In this approach, we reformulate the problem of computing tensor products in terms
of a set of Diophantine inequalities (section 4). We then look for the elementary
solutions and their relations and from these, we construct the generating
function.  We can distinguish two ways of obtaining the elementary solutions and
their relations: 

\n i) MacMahon's method which yields the generating function as an output of his
method for finding the elementary solutions and their relations (here
these are the syzygies of first, second and possibly higher orders).  This
approach is plagued by the technical difficulties of the intermediate projection
operations.  It is discussed in section 3.

\n ii) Apply the Huet's algorithm to find the elementary couplings and then
apply Grobner basis techniques to calculate a complete set of forbidden
couplings.
 This is certainly powerful and it
appears to be the most effective approach.  Grobner bases are presented in section
5 and this method is illustrated in section 4 and 6.

\n 4- {\it Diophantine equalities -- The vector basis}

In this approach, we first re-express the inequalities in terms of equalities and
then write the vector basis, relaxing the positivity constraint inherent to
the elementary couplings.
Since the basis vectors are independent the initial generating function
is simply a product of terms of the form $1/(1-E)$ for suitable monomials
$E$.
The elementary couplings are then recovered from the
projection to positive solutions (and a second
projection onto terms with integral exponents may also be necessary in
general) and the result of the projection(s) is the desired
generating function. This method and various examples are worked
out in section 7.

\subsec{Conclusion}

As it has already been stressed in the introduction, the main purpose of this
work is to prepare the ground for the analysis of fusion rules, which is the
subject of a sequel paper.  In the present work, we have reviewed the existing
techniques for computing tensor-product generating functions and presented a
comparative critical assessment of their respective virtues and limitations. In
addition, we have focused on a model formulation linking 
generating functions to  Poincar\'e series, an idea that has first been
introduced in [\Stan] and further extended in [\CCS].  Our contribution in that
respect has been to rephrase this program more explicitly, clarify some issues 
and to exemplify the procedure with many examples, some of which are new.

\vskip2cm
\centerline{\bf Acknowledgement}
We thank R.T Sharp, J. Patera and M. Walton for useful discussions on different
aspects of this work and Z. Maassarani for discussions on $osp(1,2)$.  L. B.
would also like to thank S. Lantagne and H. Roussel for computing guidance.  

\vfill\eject
\centerline{\bf REFERENCES}
\vskip 1cm
\immediate\closeout\refs \vskip 0.5cm
  \message{References}\input references
\vfill\eject

\end